\documentclass[twocolumn]{aastex701}

\hypersetup{linkcolor=red,citecolor=blue,filecolor=cyan,urlcolor=blue}

\graphicspath{{./}{figures/}}

\shorttitle{Nebular Spectra of Hydrogen-Poor Superluminous Supernovae}
\shortauthors{Blanchard et al.}

\begin{document}

\title{Hydrogen-Poor Superluminous Supernovae in the Nebular Phase: Spectral Diversity Due to Ejecta Ionization as a Probe of the Power Source}

\correspondingauthor{Peter K.~Blanchard}
\email{peter.blanchard@cfa.harvard.edu}

\author[0000-0003-0526-2248]{Peter K.~Blanchard}
\email{peter.blanchard@cfa.harvard.edu}
\affil{Center for Astrophysics \(|\) Harvard \& Smithsonian, 60 Garden St.~Cambridge, MA 02138, USA}
\affiliation{The NSF AI Institute for Artificial Intelligence and Fundamental Interactions}

\author[0000-0002-9392-9681]{Edo Berger}
\email{eberger@cfa.harvard.edu}
\affil{Center for Astrophysics \(|\) Harvard \& Smithsonian, 60 Garden St.~Cambridge, MA 02138, USA}
\affiliation{The NSF AI Institute for Artificial Intelligence and Fundamental Interactions}

\author[0000-0001-6395-6702]{Sebastian Gomez}
\email{sebastian.gomez@cfa.harvard.edu}
\affiliation{Department of Astronomy, The University of Texas at Austin, 2515 Speedway, Stop C1400, Austin, TX 78712, USA}

\author[0000-0002-2555-3192]{Matt Nicholl}
\email{matt.nicholl@qub.ac.uk}
\affiliation{Astrophysics Research Centre, School of Mathematics and Physics, Queen’s University Belfast, BT7 1NN, UK}

\author[0000-0002-7706-5668]{Ryan Chornock}
\email{chornock@berkeley.edu}
\affiliation{Department of Astronomy, University of California, Berkeley, CA 94720-3411, USA}
\affiliation{Berkeley Center for Multi-messenger Research on Astrophysical Transients and Outreach (Multi-RAPTOR), University of California, Berkeley, CA 94720-3411, USA}

\author[0000-0003-0871-4641]{Harsh Kumar}
\email{harsh.kumar@cfa.harvard.edu}
\affil{Center for Astrophysics \(|\) Harvard \& Smithsonian, 60 Garden St.~Cambridge, MA 02138, USA}
\affiliation{The NSF AI Institute for Artificial Intelligence and Fundamental Interactions}

\author[0000-0003-4768-7586]{Raffaella Margutti}
\email{rmargutti@berkeley.edu}
\affiliation{Department of Astronomy, University of California, Berkeley, CA 94720-3411, USA}
\affiliation{Berkeley Center for Multi-messenger Research on Astrophysical Transients and Outreach (Multi-RAPTOR), University of California, Berkeley, CA 94720-3411, USA}
\affiliation{Department of Physics, University of California, 366 Physics North MC 7300, Berkeley, CA 94720, USA}

\author[0000-0002-1125-9187]{Daichi Hiramatsu}
\email{dhiramatsu@ufl.edu}
\affiliation{Department of Astronomy, University of Florida, 211 Bryant Space Science Center, Gainesville, FL 32611-2055, USA}

\author[0000-0001-8023-4912]{Huei Sears}
\email{hs1349@physics.rutgers.edu}
\affiliation{Department of Physics and Astronomy, Rutgers, the State University of New Jersey, 136 Frelinghuysen Road, Piscataway, NJ 08854-8019, USA}

\begin{abstract}

We present a large sample of 39 nebular-phase optical spectra of 25 hydrogen-poor superluminous supernovae (SLSNe-I) and jointly analyze them with previously published spectra of 12 events.  We measure the properties of key emission features, namely those at 6300, 7300, and 7774 \AA\ (associated with [\ion{O}{1}], [\ion{Ca}{2}]/[\ion{O}{2}], and \ion{O}{1}, respectively), and find that SLSNe exhibit much wider spectral diversity than normal SNe Ic, primarily in the line ratio $L_{7300}/L_{6300}$, which is highly sensitive to ejecta ionization.  Some events exhibit weak [\ion{O}{1}] and a clear [\ion{O}{2}] contribution to the 7300 \AA\ feature, enhancing the ratio, along with [\ion{O}{3}] lines at 4363 and 5007 \AA.  Other SLSNe show weak or no lines of ionized oxygen.  Moreover, we find that the population exhibits decreasing $L_{7300}/L_{6300}$ over time, while a few outliers instead display sustained high or increasing ratios for extended periods.  The ratio $L_{7300}/L_{6300}$ is also correlated with the rise and decline times of the light curves, with slower events exhibiting higher ionization, the first robust connection between early light curve and late-time spectral properties, likely due to the magnetar's impact: slower-evolving SLSNe are generally powered by engines with longer spin-down timescales, which deposit more energy at later phases. Among the events with decreasing $L_{7300}/L_{6300}$, SLSNe with high ionization are on average powered by magnetars with higher thermalized spin-down power, a correlation that is most significant for events with $M_{\rm ej}\lesssim12$ M$_{\odot}$.  The ionization in the outliers with increasing $L_{7300}/L_{6300}$ may be due to late CSM interaction. $L_{7300}/L_{6300}$ and its evolution are therefore key diagnostics of SLSN engines and progenitor mass loss.      

\end{abstract}


\section{Introduction} 
\label{sec:intro}

Since their discovery, the study of hydrogen-poor superluminous supernovae (hereafter SLSNe; \citealt{Quimby2011,Chomiuk2011}) has been focused on identifying their energy sources and progenitors.  Key theories to explain their luminosities include interaction with a hydrogen-poor circumstellar medium (CSM; \citealt{ChevalierIrwin2011}), the spin-down of a millisecond magnetar \citep{KasenBildsten2010,Woosley2010}, and an overabundant production of radioactive $^{56}$Ni \citep{Gal-Yam2009}, potentially in pair instability explosions (PISNe; \citealt{HegerWoosley2002}).  Observationally, SLSNe exhibit a wide range of both timescales and luminosities (more than a factor of 10; \citealt{Chen2023ztfa,Gomez2024}), raising the possibility of diverse energy sources.  

However, among the over two hundred events discovered to date, there are none that unambiguously match PISN predictions (one event is consistent with PISN light curve predictions but has distinct spectra; \citealt{Schulze2024}).  In addition, interaction with a hydrogen-poor CSM as the dominant power source is unlikely given the low CSM densities inferred from large samples of X-ray and radio observations \citep{Margutti2018xray,Coppejans2018} and the lack of spectroscopic interaction features (narrow lines), which are seen in other SNe with strong CSM interaction (i.e., Type IIn, Ibn, Icn).  While some SLSNe have shown potential late-time signatures of interaction \citep{Yan2015,Yan2017halpha}, this interaction cannot explain their emission at peak.  Any plausible CSM model must account for the lack of narrow emission lines, for example, via an aspherical (e.g., disk-like) CSM structure.  However, detailed models of such SNe show that their peak luminosities are lower than SLSNe \citep{Suzuki2019}.         

In contrast, the magnetar scenario provides a consistent explanation for the light curves, spectra, and environments of SLSNe.  Magnetars with a range of available energy and spin-down timescales can account for their diverse luminosities and evolution timescales \citep{inserra2013,Nicholl2017,Chen2023ztfb,Gomez2024}.  Their blue early spectra, with strong \ion{O}{2} absorption, are easily explained by energy injection from a central source \citep{Dessart2012,Mazzali2016,Nicholl2017gaia16apd,Yan2017gaia16apd,Quimby2018,Aamer2025}. Furthermore, SLSNe share similar host galaxies to the engine-driven LGRBs, consistent with the expectation that metallicity likely influences the production of engine-driven transients \citep{Lunnan2014,Perley2016,Schulze2018}.  \textit{Hubble Space Telescope} observations of two SLSNe (SNe\,2015bn and 2016inl) at $\approx\!400\!-\!1100$ days after peak revealed power-law declines in their late-time light curves, as expected for SNe powered by magnetar spin-down energy \citep{Nicholl2018,Blanchard2021}.

Despite this progress in explaining the observed properties of SLSNe, several key questions remain: What are the properties of the progenitor stars that lead to magnetar formation?  How does a magnetar's energy impact the surrounding ejected layers of the exploded star and how does it get thermalized?  In addition, complex behavior such as early and late-time light curve bumps \citep{NichollSmartt2016,inserra_complexity_2017,hosseinzadeh_2021,Chen2023ztfb} and events with unusually strong iron-group absorption features \citep{blanchard2019hydrogen} may directly relate to how magnetars transfer their energy (e.g., \citealt{VurmMetzger2021,GottliebMetzger2024,Farah2025}).  Post-peak light curve bumps may also be due to interaction with low-mass CSM shells, which have been detected via light echos in several events \citep{Lunnan2018echo,Schulze2024,Gkini2025a,Gkini2025b}, providing insight into the mass-loss histories of the massive star progenitors.

Late-time studies of SLSNe, when the ejecta are optically thin and the inner ejecta are revealed, are a promising way to address the question of the magnetar's impact on the properties of the ejecta.  The presence of a newly-formed magnetar at the core of an exploding star is expected to impact the hydrodynamics and ionization/structure of the expanding ejecta and ultimately the emergent radiation.  Simulations predict the magnetar inflates a hot bubble in the innermost ejecta \citep{KasenBildsten2010,Chen2016,Chen2020magsim}.  Fluid instabilities at the bubble-ejecta boundary produce significant clumping of the ejecta \citep{Chen2016,Chen2020magsim}, which leads to unique spectroscopic signatures especially at the late nebular phase when the innermost ejecta are directly observable \citep{Jerkstrand2017,Dessart2019}.  The ejecta may also be affected by magnetar-driven outflows \citep{Metzger2015} and ionization fronts \citep{Metzger2014}.

The existing sample of nebular spectra of about 12 events shows that the inner ejecta of SLSNe are highly clumped and occasionally show enhanced ionization, key differences with normal SNe that may be due to the impact of the central engine on the inner ejecta \citep{Nicholl2016,Jerkstrand2017,inserra_complexity_2017,Dessart2019,Nicholl2019}.  The sample, however, is dominated by slowly evolving events, which are detectable for longer, preventing inferences about the population as a whole.  A much larger sample is needed to confirm this picture, or identify further complexity.  

Here we present the largest sample of nebular spectra of SLSNe studied to date, including 25 new events and 39 new spectra obtained through a dedicated five-year campaign using the Keck, Gemini, MMT, and Magellan telescopes.  We jointly analyze these spectra in combination with the 12 events studied in \citet{Nicholl2019} and uncover a large spread in the ionization state of SLSN ejecta with one dominant temporal trend consistent with magnetar powering and another suggestive of a distinct late-time ionization source such as CSM interaction.    

\begin{deluxetable*}{ccccccccc}[t]
\tablecolumns{9}
\tabcolsep0.1in\footnotesize
\tablewidth{7in}
\tablecaption{SLSN Sample, Light Curve Timescales, and Engine/Ejecta Parameters  
\label{tab:sample}}
\tablehead {
\colhead {SN} &
\colhead {$z$}   &
\colhead {$r$-band peak\tablenotemark{*}}     &
\colhead {$t_{\rm r}$ (days)\tablenotemark{*,$\dagger$}} &
\colhead {$t_{\rm d}$ (days)\tablenotemark{*,$\ddagger$}} &
\colhead {$M_{\rm ej}$\tablenotemark{*}} &
\colhead {log($E_{\rm K}$/erg)\tablenotemark{*}} &
\colhead {$P$\tablenotemark{*}} &
\colhead {log($B$/G)\tablenotemark{*}} \\
\colhead {} &
\colhead {} &
\colhead {(MJD)} &
\colhead {(days)} &
\colhead {(days)} &
\colhead {(M$_{\odot}$)} &
\colhead {} &
\colhead {(ms)} &
\colhead {}
}   
\startdata
SN\,2018fd & 0.263 & 58084 & 124.5$^{+25.6}_{-31.8}$ & 150.0$^{+11.1}_{-7.7}$ & 38.2$^{+20.4}_{-16.9}$ & 51.38$^{+0.15}_{-0.17}$ & 3.40$^{+0.66}_{-0.62}$ & 13.71$^{+0.15}_{-0.12}$ \\
SN\,2018avk & 0.132 & 58269 & 68.0$^{+2.2}_{-4.9}$ & 80.6$^{+10.9}_{-11.1}$ & 8.5$^{+5.6}_{-2.9}$ & 51.22$^{+0.25}_{-0.18}$ & 3.89$^{+1.88}_{-0.62}$ & 13.65$^{+0.22}_{-0.16}$ \\
SN\,2018bym & 0.274 & 58276 & 41.7$^{+0.7}_{-0.7}$ & 39.1$^{+0.3}_{-0.6}$ & 3.1$^{+1.1}_{-0.5}$ & 51.13$^{+0.13}_{-0.08}$ & 2.40$^{+0.27}_{-0.29}$ & 13.50$^{+0.16}_{-0.11}$ \\
SN\,2018gft & 0.232 & 58428 & 76.2$^{+0.9}_{-1.0}$ & 67.4$^{+0.6}_{-0.8}$ & 7.7$^{+0.4}_{-0.3}$ & 51.27$^{+0.02}_{-0.01}$ & 1.86$^{+0.18}_{-0.19}$ & 13.88$^{+0.12}_{-0.10}$ \\
SN\,2018hti & 0.061 & 58470 & 55.2$^{+0.3}_{-1.1}$ & 52.5$^{+1.0}_{-0.6}$ & 15.1$^{+5.0}_{-2.9}$ & 51.50$^{+0.11}_{-0.10}$ & 2.24$^{+0.24}_{-0.20}$ & 13.80$^{+0.09}_{-0.09}$ \\
SN\,2019neq & 0.106 & 58733 & 28.8$^{+0.5}_{-0.1}$ & 25.8$^{+0.1}_{-0.6}$ & 1.7$^{+0.4}_{-0.5}$ & 51.12$^{+0.11}_{-0.14}$ & 3.25$^{+0.37}_{-0.26}$ & 13.72$^{+0.16}_{-0.13}$ \\
SN\,2020auv & 0.280 & 58876 & 22.8$^{+2.0}_{-3.7}$ & 29.3$^{+3.1}_{-3.0}$ & 23.0$^{+10.5}_{-7.0}$ & 52.06$^{+0.18}_{-0.26}$ & 1.00$^{+0.49}_{-0.23}$ & 14.66$^{+0.14}_{-0.13}$ \\
SN\,2020onb & 0.153 & 59079 & 46.8$^{+1.3}_{-1.8}$ & 46.2$^{+2.1}_{-2.3}$ & 5.1$^{+1.6}_{-1.2}$ & 51.42$^{+0.09}_{-0.16}$ & 4.72$^{+0.50}_{-0.64}$ & 14.18$^{+0.18}_{-0.12}$ \\
SN\,2020tcw & 0.064 & 59135 & 38.6$^{+1.1}_{-3.5}$ & 34.8$^{+2.9}_{-4.5}$ & 7.1$^{+9.0}_{-3.1}$ & 51.40$^{+0.21}_{-0.24}$ & 3.36$^{+1.90}_{-1.57}$ & 14.50$^{+0.17}_{-0.15}$ \\
SN\,2020qlb & 0.158 & 59149 & 86.6$^{+1.9}_{-7.7}$ & 102.5$^{+1.5}_{-13.1}$ & 48.4$^{+1.9}_{-12.2}$ & 52.29$^{+0.05}_{-0.26}$ & 0.94$^{+1.67}_{-0.16}$ & 14.07$^{+0.10}_{-0.22}$ \\
SN\,2020rmv & 0.262 & 59117 & 57.8$^{+2.6}_{-1.1}$ & 95.8$^{+3.6}_{-2.8}$ & 32.1$^{+7.1}_{-5.1}$ & 51.44$^{+0.09}_{-0.09}$ & 5.25$^{+0.72}_{-0.72}$ & 14.16$^{+0.27}_{-0.22}$ \\
SN\,2020wnt & 0.032 & 59208 & 72.6$^{+1.2}_{-1.2}$ & 67.7$^{+1.2}_{-1.0}$ & 4.7$^{+0.8}_{-0.6}$ & 51.00$^{+0.06}_{-0.07}$ & 4.32$^{+0.48}_{-0.48}$ & 13.60$^{+0.16}_{-0.12}$ \\
SN\,2020znr & 0.114 & 59235 & 72.8$^{+0.9}_{-1.2}$ & 112.2$^{+2.2}_{-3.1}$ & 1.0$^{+0.3}_{-0.1}$ & 49.52$^{+0.09}_{-0.09}$ & 2.94$^{+0.22}_{-0.38}$ & 13.52$^{+0.08}_{-0.10}$ \\
SN\,2020adkm & 0.226 & 59245 & 73.7$^{+4.0}_{-3.4}$ & 83.7$^{+6.1}_{-6.1}$ & 62.1$^{+20.7}_{-19.2}$ & 51.61$^{+0.15}_{-0.15}$ & 1.15$^{+0.35}_{-0.29}$ & 13.85$^{+0.12}_{-0.11}$ \\
SN\,2020aewh & 0.345 & 59258 & 65.4$^{+7.8}_{-8.8}$ & 84.0$^{+6.3}_{-6.5}$ & 20.0$^{+9.8}_{-6.4}$ & 51.78$^{+0.24}_{-0.32}$ & 1.49$^{+0.67}_{-0.50}$ & 13.97$^{+0.12}_{-0.16}$ \\
SN\,2020abjc & 0.219 & 59260 & 131.8$^{+7.4}_{-8.8}$ & 161.2$^{+6.1}_{-7.9}$ & 31.6$^{+6.7}_{-6.9}$ & 51.14$^{+0.13}_{-0.12}$ & 4.53$^{+0.49}_{-0.60}$ & 13.48$^{+0.13}_{-0.14}$ \\
SN\,2022le & 0.249 & 59707 & 198.6$^{+17.1}_{-14.2}$ & 131.8$^{+41.5}_{-34.2}$ & 0.2$^{+0.4}_{-0.1}$ & 48.14$^{+0.42}_{-0.25}$ & 2.38$^{+0.19}_{-0.22}$ & 13.49$^{+0.41}_{-0.22}$ \\
SN\,2022acsx & 0.279 & 59960 & 47.6$^{+4.6}_{-2.5}$ & 43.7$^{+4.3}_{-3.1}$ & 16.7$^{+13.3}_{-6.7}$ & 51.49$^{+0.18}_{-0.16}$ & 1.78$^{+0.46}_{-0.50}$ & 14.01$^{+0.22}_{-0.15}$ \\
SN\,2023cmx & 0.233 & 60055 & 81.7$^{+2.8}_{-2.5}$ & 76.7$^{+3.3}_{-3.4}$ & 7.8$^{+1.6}_{-1.2}$ & 51.40$^{+0.10}_{-0.09}$ & 1.63$^{+0.25}_{-0.20}$ & 13.17$^{+0.20}_{-0.18}$ \\
SN\,2023wwx & 0.308 & 60307 & 88.2$^{+6.7}_{-9.8}$ & 96.9$^{+6.3}_{-6.1}$ & 24.8$^{+22.9}_{-10.8}$ & 51.55$^{+0.36}_{-0.19}$ & 1.60$^{+0.63}_{-0.60}$ & 13.80$^{+0.18}_{-0.16}$ \\
SN\,2024amf & 0.068 & 60387 & 71.8$^{+1.1}_{-1.1}$ & 61.7$^{+1.8}_{-1.5}$ & 5.2$^{+1.2}_{-0.8}$ & 50.87$^{+0.11}_{-0.12}$ & 6.33$^{+0.60}_{-0.60}$ & 14.27$^{+0.49}_{-0.21}$ \\
SN\,2024ahr & 0.086 & 60413 & 97.1$^{+5.7}_{-5.9}$ & 92.7$^{+1.8}_{-2.1}$ & 9.9$^{+1.1}_{-1.3}$ & 51.27$^{+0.07}_{-0.11}$ & 3.21$^{+0.35}_{-0.24}$ & 13.53$^{+0.38}_{-0.15}$ \\
SN\,2024dde & 0.054 & 60425 & 72.9$^{+1.6}_{-1.8}$ & 67.2$^{+3.1}_{-3.7}$ & 14.9$^{+9.1}_{-5.8}$ & 51.11$^{+0.22}_{-0.20}$ & 2.90$^{+0.97}_{-1.21}$ & 14.34$^{+0.19}_{-0.15}$ \\
SN\,2024jlc & 0.039 & 60506 & 53.6$^{+2.0}_{-2.0}$ & 54.2$^{+6.0}_{-4.2}$ & 9.3$^{+3.8}_{-2.8}$ & 51.08$^{+0.18}_{-0.15}$ & 4.06$^{+3.78}_{-1.81}$ & 14.70$^{+0.20}_{-0.19}$ \\
SN\,2024jwx & 0.120 & 60509 & 61.1$^{+1.4}_{-0.9}$ & 50.9$^{+5.1}_{-4.7}$ & 3.4$^{+2.2}_{-1.1}$ & 50.35$^{+0.30}_{-0.26}$ & 3.55$^{+0.85}_{-0.60}$ & 13.93$^{+0.21}_{-0.13}$ \\
\enddata
\tablenotetext{*}{Values for SN\,2022acsx and later events from this work; earlier events from \citet{Gomez2024}.}
\tablenotetext{\dagger}{Rise time, defined as the rest-frame days from explosion to $r$-band peak.}
\tablenotetext{\ddagger}{Decline time, defined as the rest-frame days since bolometric peak for the luminosity to decline by a factor of $e$.}
\end{deluxetable*}

The paper is organized as follows.  In Section \ref{sec:sample} we describe the sample and spectroscopic observations.  In Section \ref{sec:specprop} we measure and analyze the properties of key emission lines and ratios.  In Section \ref{sec:LCmag_comp} we test for correlations between the key ionization-sensitive ratio $L_{7300}/L_{6300}$ and light curve timescales and magnetar properties.  In Section \ref{sec:disc} we compare our observed nebular properties of SLSNe with normal SNe Ic and discuss the likely sources of the ejecta ionization.  We summarize and conclude in Section \ref{sec:conc}.

\section{SLSN Sample and Spectroscopic Observations}
\label{sec:sample}

\begin{figure*}
\centering
\includegraphics[scale=0.5]{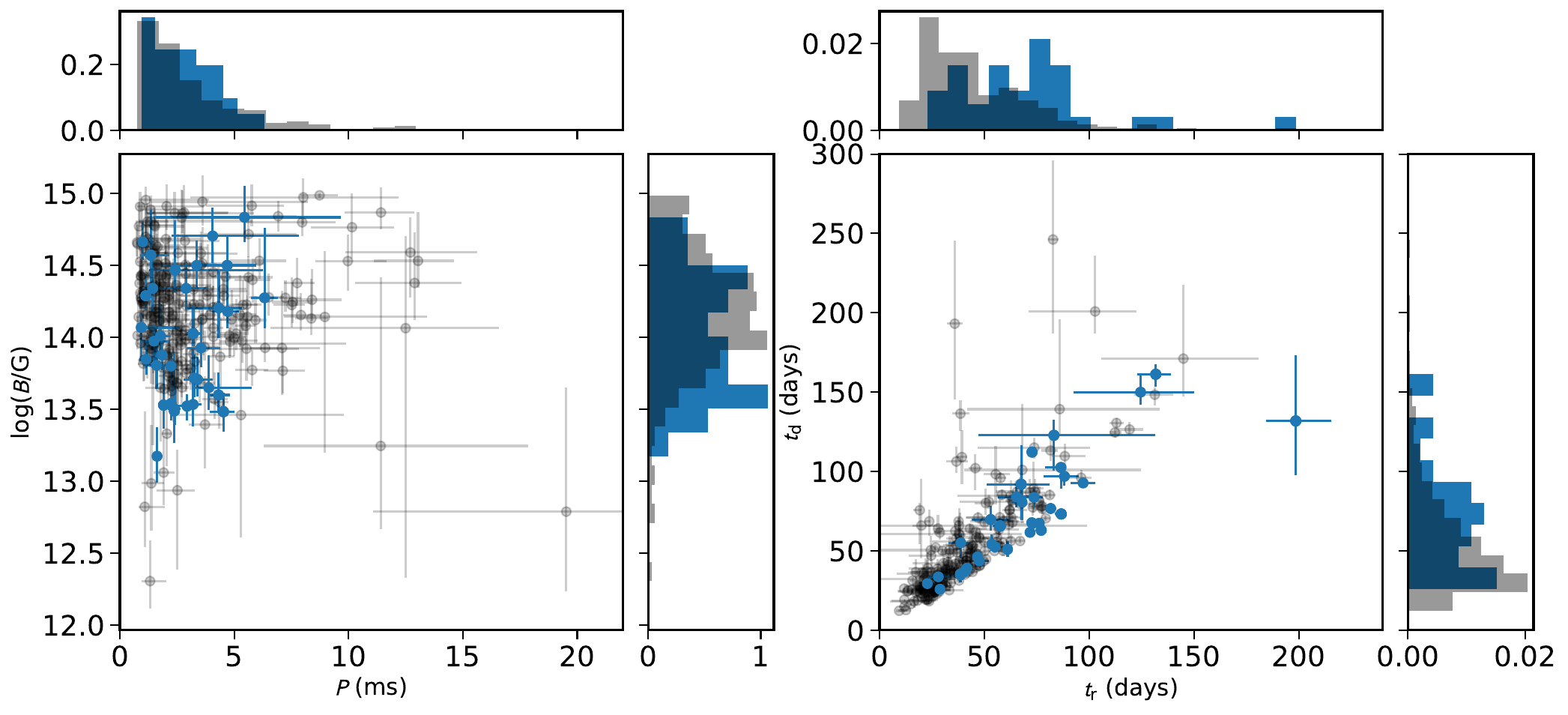}
\caption{Comparison of our nebular sample (blue) magnetar parameters (left) and light curve rise and decline timescales (right) with those for the full SLSN light curve sample (gray; excluding bronze events) from \citet{Gomez2024}.  The nebular sample is representative of nearly the full range of properties, though with a slight over-representation of slowly evolving SLSNe.}
\label{fig:paramcomp}
\end{figure*}

Our sample (shown in Table \ref{tab:sample}) consists of 25 SLSNe discovered in the years $2018-2024$.  We include all events from our follow-up program with spectroscopic observations beyond $\approx$150 rest-frame days after peak, the rough time SLSNe are expected to transition into the nebular phase.  We do not impose a strict luminosity cut, but instead include all objects that are spectroscopically consistent with the SLSN population \citep{Gomez2024}.  To assess how our late-time sample compares with the population, in Figure \ref{fig:paramcomp} we show the inferred engine parameters (initial spin, $P$, and magnetic field strength, $B$) and measured light curve timescales (rise time, $t_{\rm r}$, defined as the rest-frame days from explosion to $r$-band peak, and decline time, $t_{\rm d}$, the $e$-folding time since bolometric peak) of our nebular sample with those for the full SLSN sample.  The SLSNe with nebular spectra span nearly the full range of parameters, with a slight over representation of slowly evolving events as expected.  Importantly, our sample adds nebular spectra of 10 new events with $t_{\rm d}\lesssim60$ days.

We obtained spectroscopic observations of our sample using Binospec \citep{Fabricant2019} mounted on the MMT 6.5-m telescope, the Low Resolution Imaging Spectrometer (LRIS; \citealt{Oke1995}) mounted on the Keck I 10-m telescope, the Deep Imaging Multi-Object Spectrograph (DEIMOS; \citealt{Faber2003}) mounted on the Keck II 10-m telescope, the Gemini Multi Object Spectrograph (GMOS; \citealt{Hook2004}) on the Gemini North and South 8-m telescopes, and the Inamori-Magellan Areal Camera and Spectrograph (IMACS; \citealt{Dressler2011}) on the 6.5-m Magellan Baade telescope.  

\begin{figure*}[t!]
\centering
\includegraphics[scale=0.6]{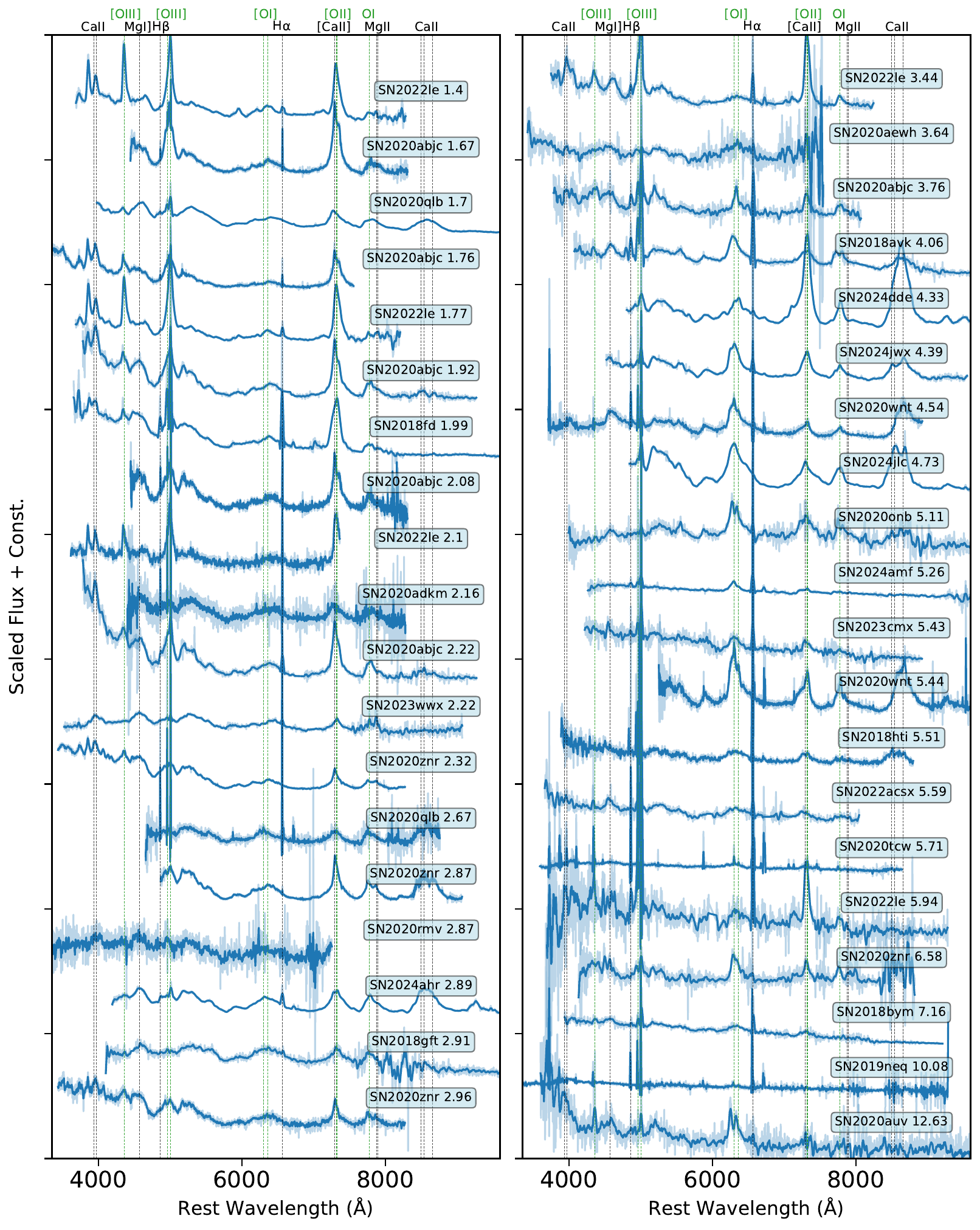}
\caption{Spectra of our SLSN sample in order of normalized phase (rest-frame days since peak normalized by the characteristic decline time $t_{\rm d}$).  Original and smoothed spectra are shown.  Temporal differences are clearly apparent, with spectra at $t/t_{\rm d}\lesssim3$ exhibiting broad, underdeveloped lines at 6300 \AA\ (the location of [\ion{O}{1}]) with often strong emission at 7300 \AA\ (the location of [\ion{Ca}{2}] and [\ion{O}{2}]).  A subset of these show obvious [\ion{O}{3}] lines at 4363 and 5007 \AA\ (SN\,2020abjc, SN\,2022le).  At later phases, the 6300 \AA\ emission is narrower and comparable to or stronger than the emission at 7300 \AA.  In addition, emission near the \ion{O}{1} recombination line at 7774 \AA\ is present in events with sufficient coverage and S/N; this emission exhibits a red shoulder in several cases.}
\label{fig:allspec}
\end{figure*}

Details of the observations are presented in Table~\ref{tab:spec}.  We extracted 1D wavelength-calibrated spectra and used observations of standard stars for relative flux calibration.  We used the python code {\tt Pypeit} \citep{Pypeit} to reduce our Gemini and Keck observations and standard routines in {\tt IRAF} \citep{Tody1986,Tody1993} using {\tt redspec} \citep{redspec2024} to reduce our Binospec and IMACS observations.  All spectra are corrected for Galactic extinction and shown in Figure \ref{fig:allspec}.

\begin{deluxetable*}{cccccc}[t!]
\tablecolumns{6}
\tabcolsep0.1in\footnotesize
\tablewidth{7in}
\tablecaption{Spectroscopic Observations  
\label{tab:spec}}
\tablehead {
\colhead {SN} &
\colhead {Date (UT)}   &
\colhead {MJD}     &
\colhead {Phase\tablenotemark{*}} &
\colhead {Telescope} &
\colhead{Instrument\tablenotemark{$\dagger$}}             
}   
\startdata
SN\,2018fd & 2018-12-09 & 58461 & 299 & Gemini & GMOS \\
SN\,2018avk & 2019-06-05 & 58640 & 327 & Gemini & GMOS \\
SN\,2018bym & 2019-05-30 & 58633 & 280 & Gemini & GMOS \\
SN\,2018gft & 2019-07-05 & 58670 & 196 & Gemini & GMOS \\
SN\,2018hti & 2019-10-20 & 58776 & 289 & Magellan & IMACS \\
SN\,2019neq & 2020-06-20 & 59020 & 260 & Keck & LRIS \\
SN\,2020auv & 2021-05-16 & 59351 & 371 & Gemini & GMOS \\
SN\,2020onb & 2021-05-17 & 59352 & 236 & Gemini & GMOS \\
SN\,2020tcw & 2021-05-11 & 59346 & 199 & MMT & Binospec \\
SN\,2020qlb & 2021-05-17 & 59352 & 175 & Gemini & GMOS \\
            & 2021-09-09 & 59466 & 273 & Keck & DEIMOS \\
SN\,2020rmv & 2021-09-09 & 59467 & 275 & MMT & Binospec \\
SN\,2020wnt & 2021-11-07 & 59526 & 307 & MMT & Binospec \\
            & 2022-01-09 & 59588 & 368 & Keck & DEIMOS \\
SN\,2020znr & 2021-11-03 & 59522 & 261 & MMT & Binospec \\
            & 2022-01-09 & 59588 & 322 & Keck & DEIMOS \\
            & 2022-01-21 & 59600 & 333 & MMT & Binospec \\
            & 2023-04-12 & 60047 & 738 & Gemini & GMOS \\
SN\,2020adkm & 2021-09-09 & 59466 & 180 & Keck & DEIMOS \\
SN\,2020aewh & 2022-03-31 & 59670 & 306 & Gemini & GMOS \\
SN\,2020abjc & 2022-01-09 & 59589 & 270 & Keck & DEIMOS \\
             & 2022-01-26 & 59606 & 284 & MMT & Binospec \\
             & 2022-02-26 & 59636 & 309 & Gemini & GMOS \\
             & 2022-03-31 & 59669 & 336 & Keck & DEIMOS \\
             & 2022-04-26 & 59695 & 357 & Gemini & GMOS \\
             & 2023-02-23 & 59999 & 606 & Gemini & GMOS \\
SN\,2022le & 2022-12-25 & 59938 & 185 & Gemini & GMOS \\
           & 2023-02-23 & 59999 & 233 & Gemini & GMOS \\
           & 2023-04-18 & 60052 & 276 & MMT & Binospec \\
           & 2023-11-25 & 60273 & 453 & Gemini & GMOS \\
           & 2025-01-09 & 60685 & 783 & Gemini & GMOS \\
SN\,2022acsx & 2023-11-24 & 60272 & 244 & Gemini & GMOS \\
SN\,2023cmx  & 2024-09-15 & 60569 & 416 & Gemini & GMOS \\
SN\,2023wwx  & 2024-10-05 & 60589 & 215 & Gemini & GMOS \\
SN\,2024amf  & 2025-02-27 & 60734 & 324 & Gemini & GMOS \\
SN\,2024ahr  & 2025-01-29 & 60705 & 268 & Gemini & GMOS \\
SN\,2024dde  & 2025-02-25 & 60732 & 291 & Gemini & GMOS \\
SN\,2024jlc  & 2025-04-06 & 60772 & 256 & Gemini & GMOS \\
SN\,2024jwx  & 2025-03-25 & 60760 & 224 & Gemini & GMOS \\
\enddata
\tablenotetext{*}{Rest-frame phase relative to $r$-band peak.}
\tablenotetext{\dagger}{Resolutions for the setups used are $\approx 14,7,7,3,4$ \AA\ for GMOS, IMACS, LRIS, DEIMOS, and Binospec, respectively.}
\end{deluxetable*}

\begin{figure*}[t]
\centering
\includegraphics[width=\linewidth]{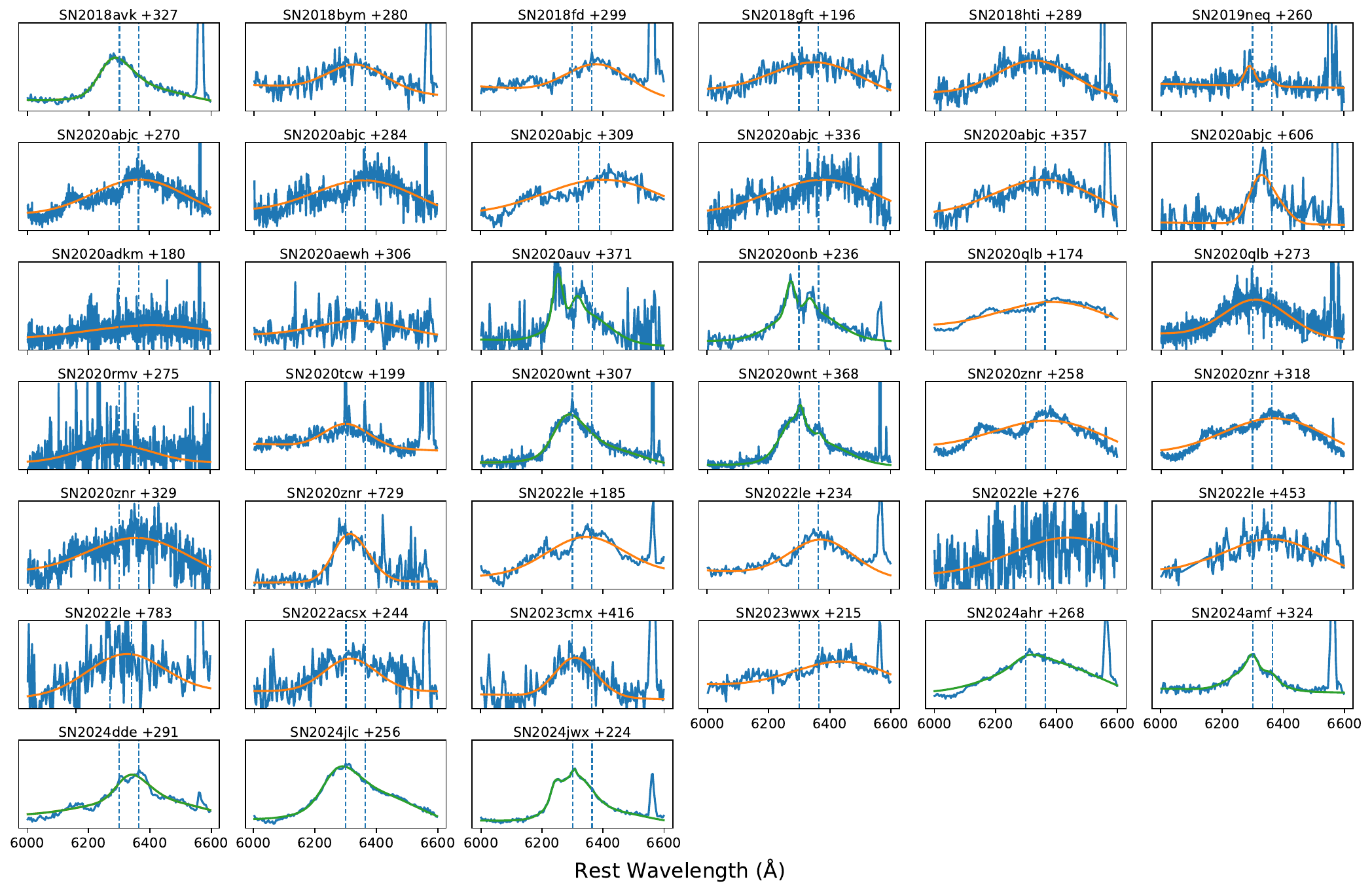}
\caption{Spectra of our sample zoomed-in on the 6300 \AA\ feature with the single (orange) or multi (green) component Gaussian fits to the line profiles shown.  The location of the [\ion{O}{1}] $\lambda6300$ doublet is marked with vertical dashed lines.  The phases correspond to rest-frame days since peak.}
\label{fig:OI}
\end{figure*}

\begin{figure}[t]
\centering
\includegraphics[width=\linewidth]{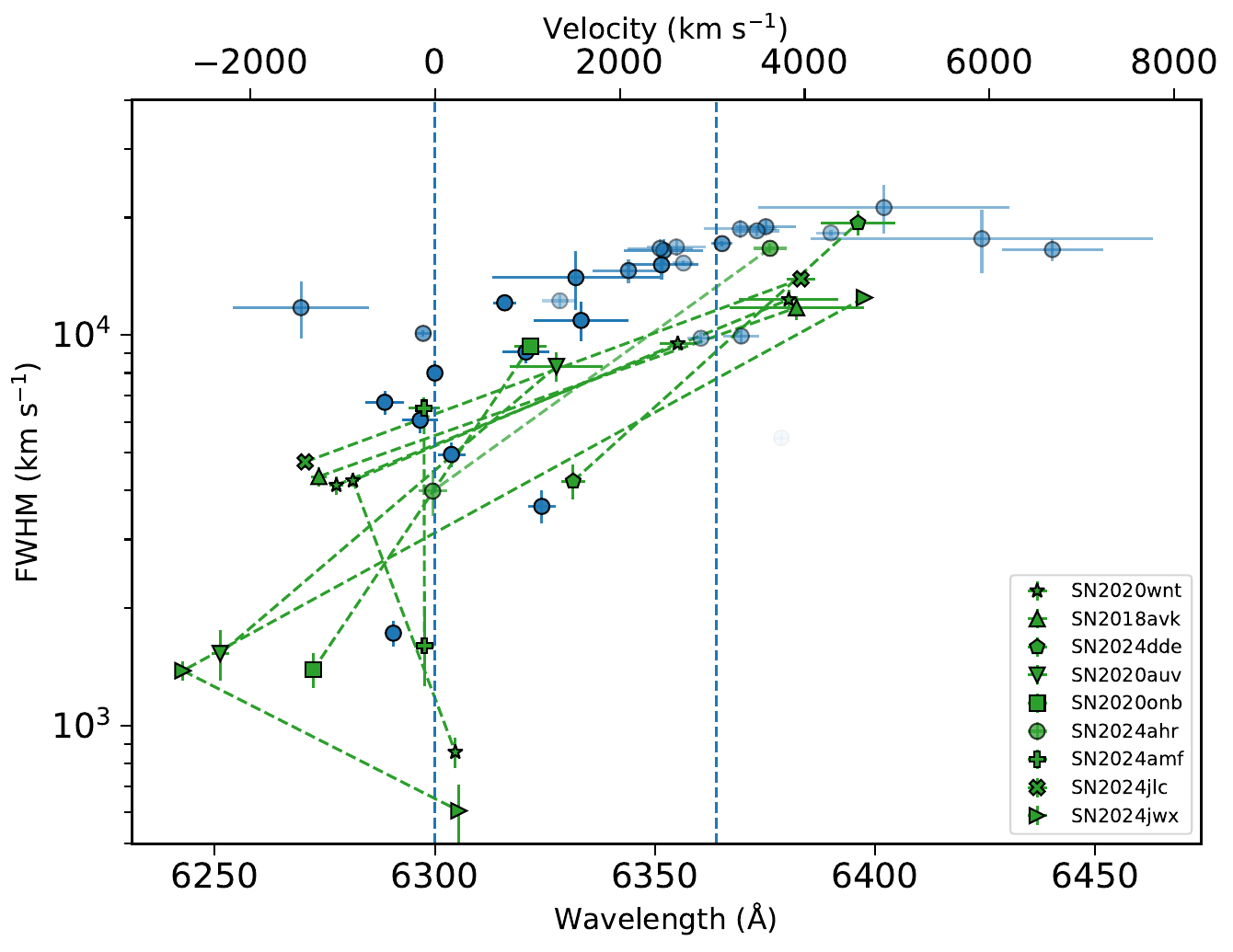}
\caption{Line width (FWHM) versus central wavelength for the fits shown in Figure \ref{fig:OI} with the location of the [\ion{O}{1}] $\lambda6300$ doublet marked (blue dashed lines).  The green markers are profiles for which the best fit consists of two or three components (the values for each component are connected by a dashed line).  The transparency of the points is proportional to the normalized phase of the spectra, with lighter shades corresponding to $t/t_{\rm d}<4$ and solid points to $t/t_{\rm d}>4$.  A temporal trend is apparent where spectra taken at earlier phases have broad profiles, corresponding to high velocities, that are systematically centered redward of [\ion{O}{1}].  Several events with lower velocities have components with an appreciable blueshift.}  
\label{fig:OIvel}
\end{figure}

\begin{figure*}[t]
\centering
\includegraphics[width=\linewidth]{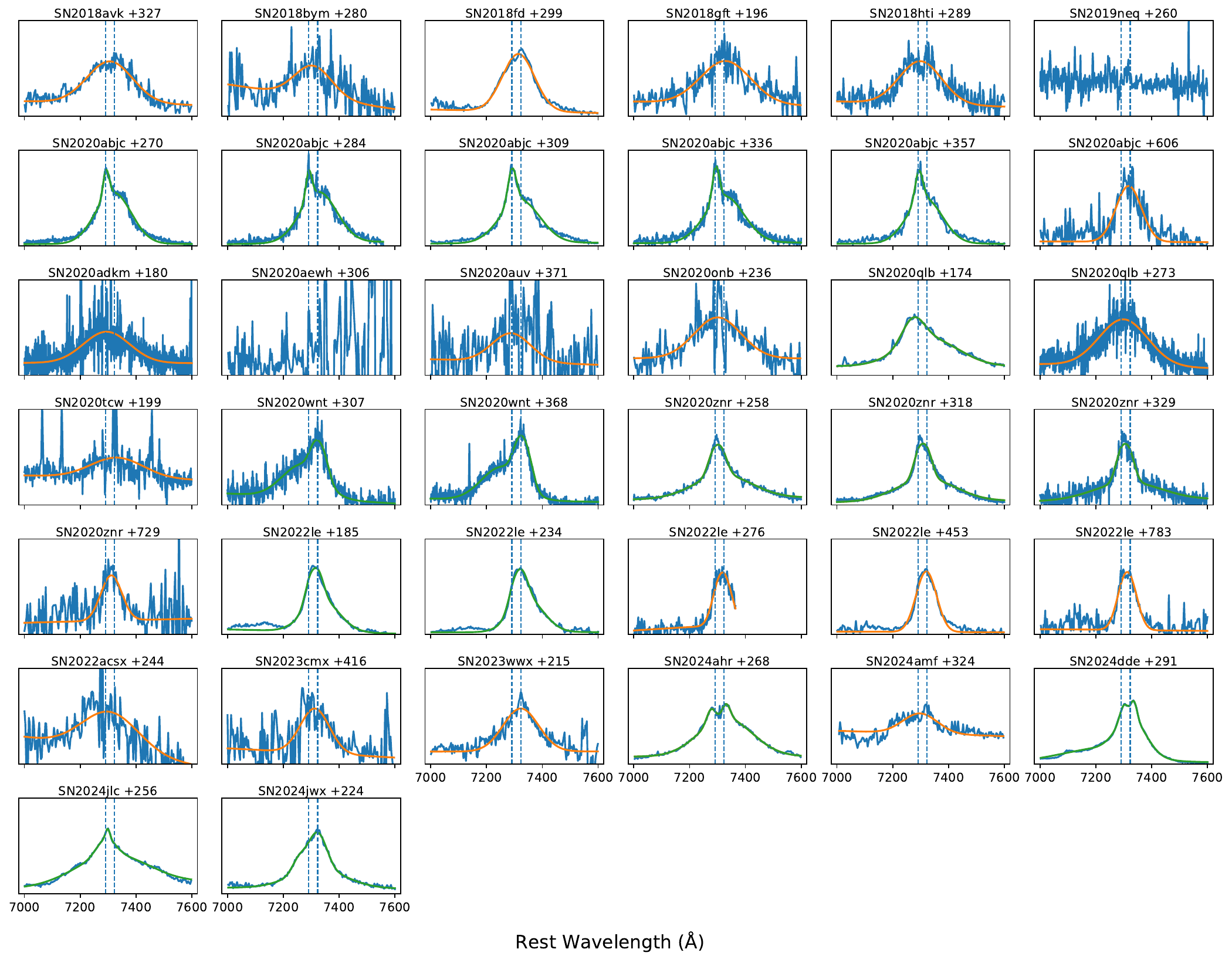}
\caption{Same as Figure \ref{fig:OI} but for the 7300 \AA\ feature.  The location of the [\ion{Ca}{2}] $\lambda7300$ doublet is marked with vertical dashed lines.  The profiles that are best fit with two components (green) mostly show excess emission redward of [\ion{Ca}{2}] with the exception of SN\,2020wnt which has a prominent component blueward.}   
\label{fig:CaII}
\end{figure*}

\begin{figure}[t]
\centering
\includegraphics[width=\linewidth]{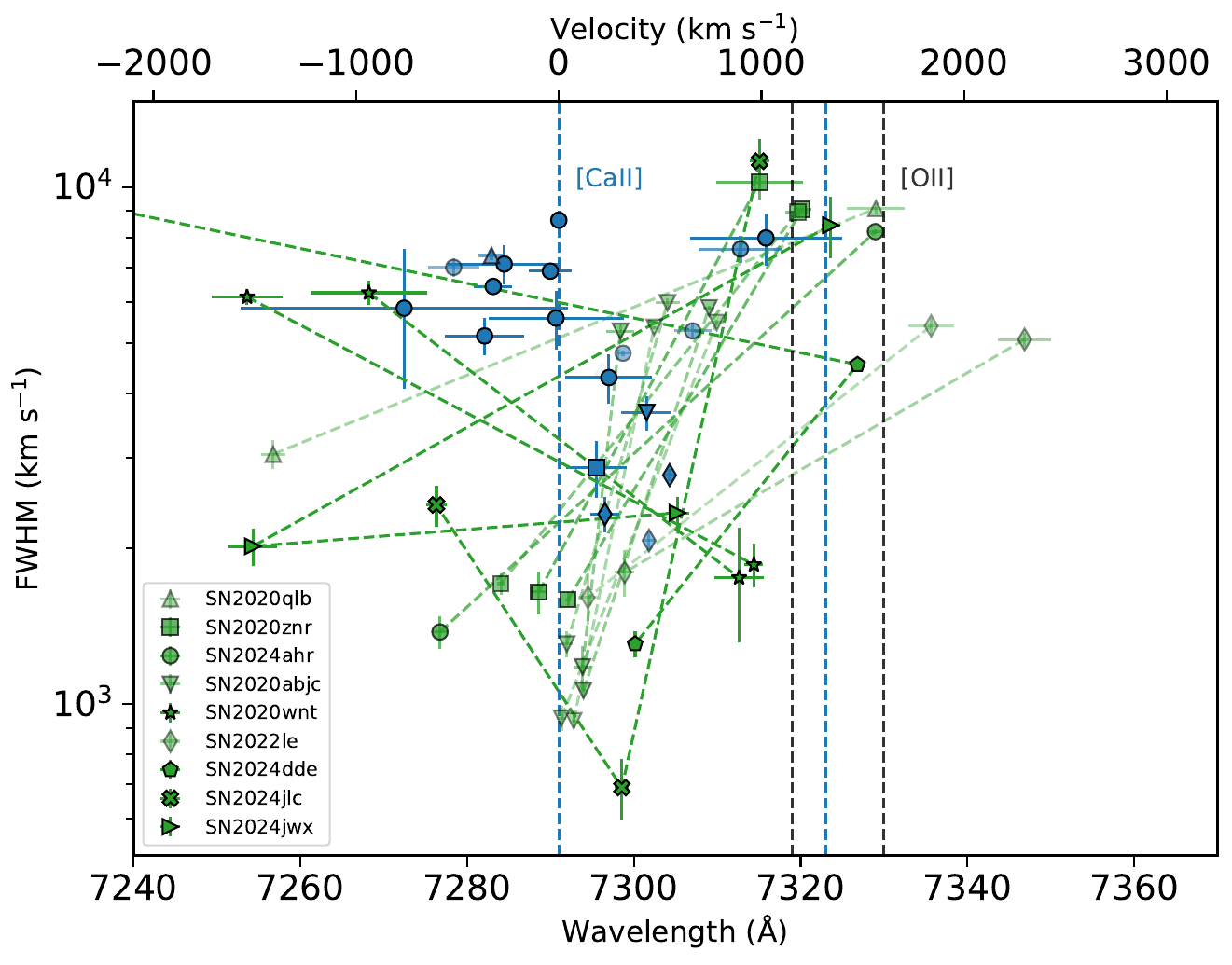}
\caption{Same as Figure \ref{fig:OIvel} but for the 7300 \AA\ feature fits in Figure \ref{fig:CaII} with the locations of the [\ion{Ca}{2}] and [\ion{O}{2}] doublets marked (blue and black dashed lines, respectively).  Several of the single component profiles are shifted redward of [\ion{Ca}{2}] and all of the profiles with two or more components have one component shifted to the red.}
\label{fig:CaIIvel}
\end{figure}

\begin{figure*}[t]
\centering
\includegraphics[width=\linewidth]{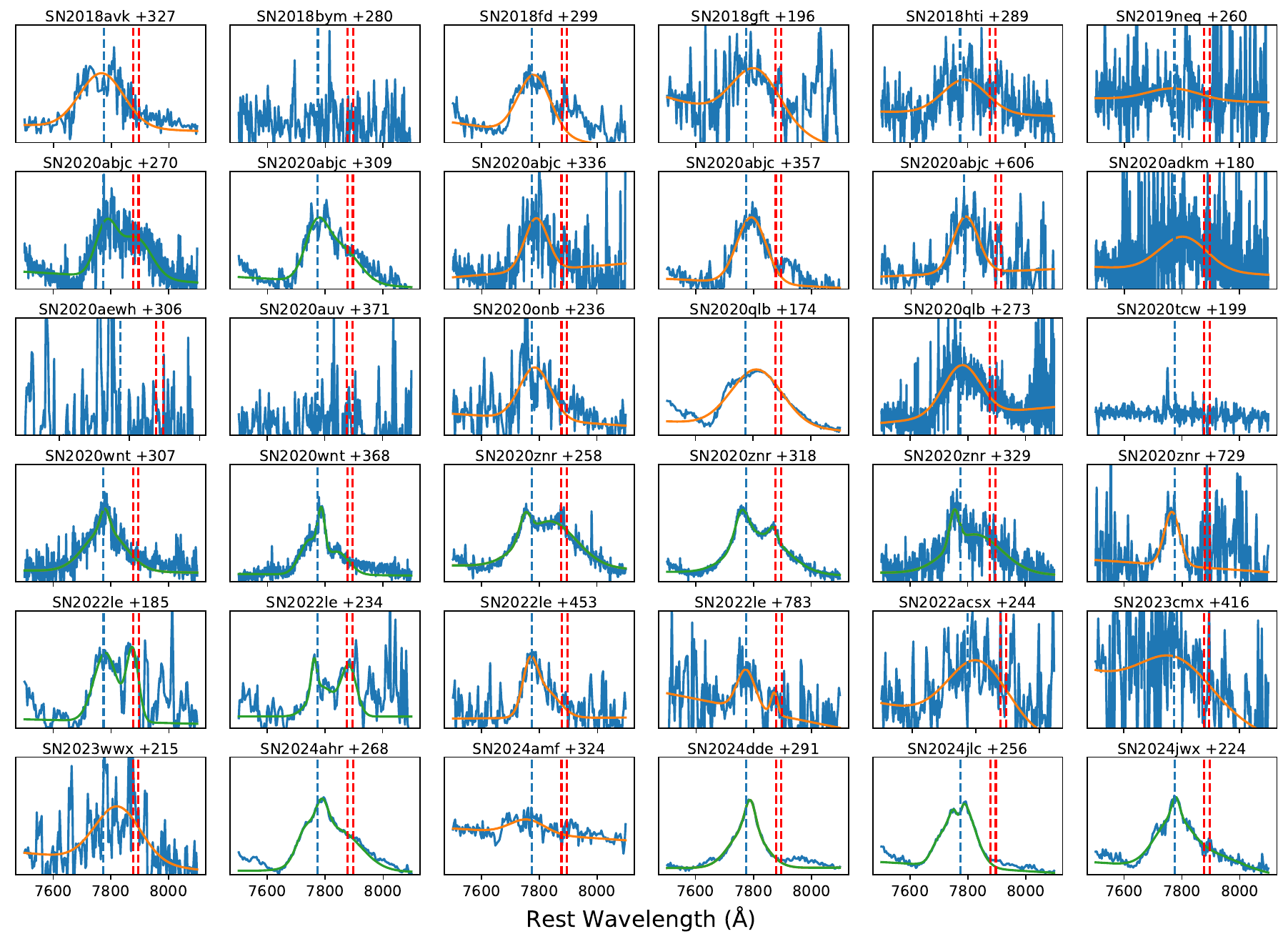}
\caption{Same as Figure \ref{fig:OI} but for the 7774 \AA\ feature.  The locations of \ion{O}{1} $\lambda7774$ (dashed blue line) and \ion{Mg}{2} $\lambda$7877,7896 (dashed red lines) are marked.  The complexity of this spectral feature varies considerably, with some events exhibiting prominent emission redward of \ion{O}{1} likely due to \ion{Mg}{2}.}
\label{fig:OI7774}
\end{figure*}

\begin{figure}[t]
\centering
\includegraphics[width=\linewidth]{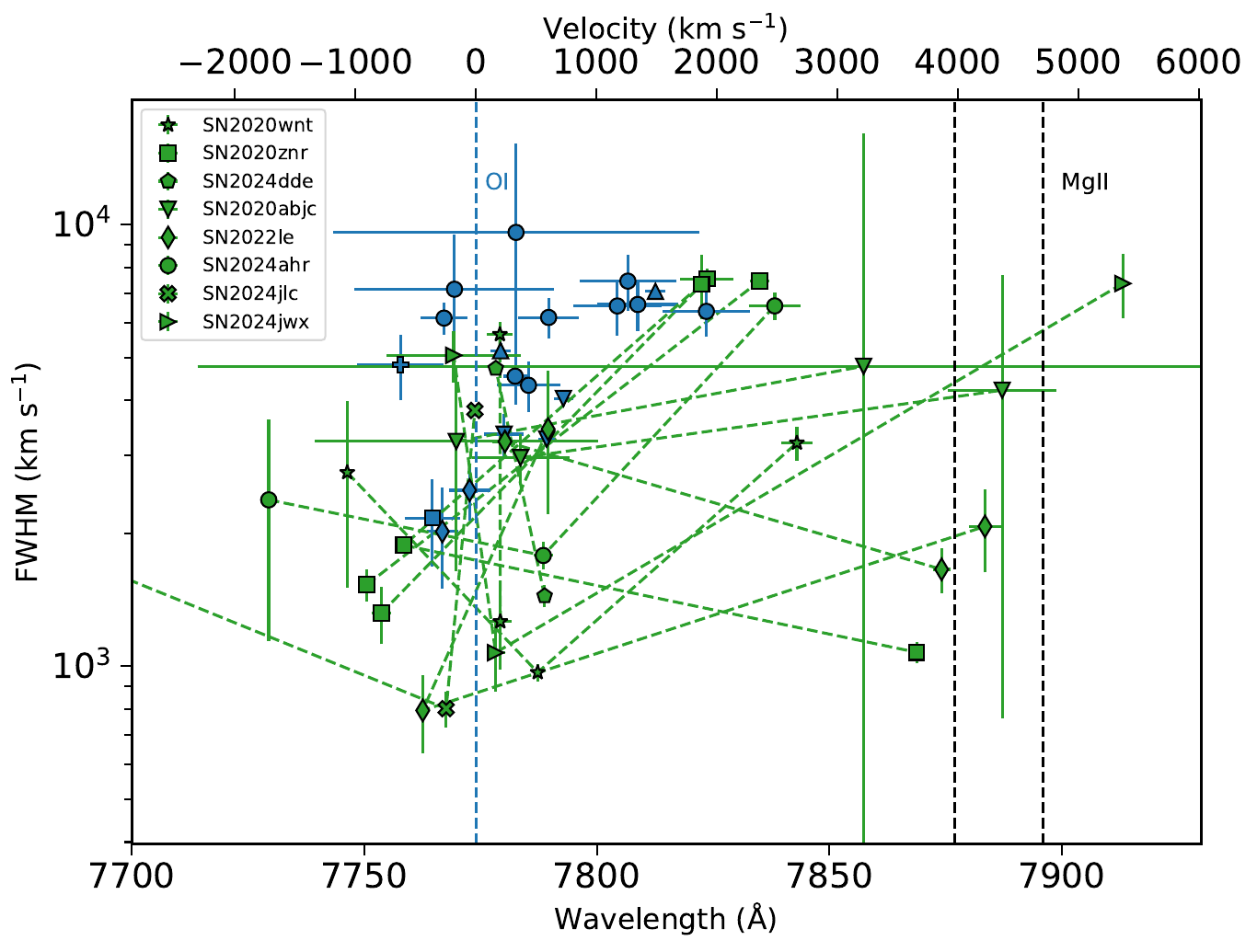}
\caption{Same as Figure \ref{fig:OIvel} but for the 7774 \AA\ feature fits in Figure \ref{fig:OI7774} with the locations of \ion{O}{1} $\lambda7774$ (dashed blue line) and \ion{Mg}{2} $\lambda$7877,7896  (dashed black lines) marked.  Most of the profiles best fit with multiple components have one component shifted redward of \ion{O}{1}, with some centered near \ion{Mg}{2}.  In addition, several events have \ion{O}{1}/\ion{Mg}{2} components with very low velocities of $\approx800-2000$ km s$^{-1}$.}
\label{fig:OI7774vel}
\end{figure}

\section{Nebular Emission Line Properties}
\label{sec:specprop}

Similar to normal SESNe, SLSNe are known to show strong nebular emission lines from [\ion{O}{1}] $\lambda6300$, [\ion{Ca}{2}] $\lambda7300$, and \ion{Mg}{1}] $\lambda4571$ \citep{Jerkstrand2017,Nicholl2019}.  In addition to these common features, SLSNe show emission from \ion{O}{1} $\lambda7774$, which is rarely seen in normal SESNe, as well as overall higher flux in the region $4000-5000$ \AA\ compared to normal SESNe, which may be due to many overlapping lines from [\ion{Fe}{2}].  The events in our sample show all of these general features (Figure \ref{fig:allspec}), though our larger sample enables a more detailed exploration of the range of properties.  

The spectra shown in Figure \ref{fig:allspec} exhibit striking diversity with a wide range of emission line strengths at 6300 and 7300 \AA, the features often attributed to [\ion{O}{1}] $\lambda6300$ and [\ion{Ca}{2}] $\lambda7300$.  Some events have strong emission at 7300 \AA\ but weak emission at 6300 \AA, with a few showing obvious [\ion{O}{3}] $\lambda$5007 emission (SN\,2020abjc and SN\,2022le).  In addition, events with sufficient coverage and signal-to-noise show emission near 7774 \AA, the location of a key \ion{O}{1} recombination line.     

\subsection{Line Profiles, Shifts, and Widths}

To understand this spectral diversity, we begin by measuring the profiles, line centers, and widths of the prominent nebular emission features at 6300, 7300, and 7774 \AA.  For all lines, we fit one or more Gaussian components.  We estimate the background continuum level by fitting a line to regions free of emission features on the red and blue sides of each emission line.  For the doublet lines ([\ion{O}{1}] $\lambda$6300, [\ion{Ca}{2}] $\lambda$7300), each primary Gaussian component itself consists of two Gaussians representing the two lines of the doublet with fixed wavelength separations and appropriate flux ratios assuming the lines are optically thin (i.e.~3-to-1 for the flux ratio of the 6300 \AA\ line relative to 6364 \AA; \citealt{OsterbrockFerland2006}).  

For each SN and emission line, we start by fitting one Gaussian component.  If clear patterns are present in the residuals, we progressively add additional Gaussian components until random residuals are achieved.       

\subsubsection{6300 \AA\ Feature}
\label{sec:6300}

In Figure \ref{fig:OI} we show our fits to the 6300 \AA\ feature, often attributed to the [\ion{O}{1}] $\lambda$6300 doublet, one of the strongest lines seen in nebular spectra of stripped-envelope SNe of any sub-type and therefore a key diagnostic used in the literature (e.g., \citealt{Matheson2001,Taubenberger2009,Milisavljevic2010,Jerkstrand2017,Nicholl2019,Fang2022}).  While many SLSNe in our sample have [\ion{O}{1}] $\lambda$6300 lines that are well-represented by a single broad Gaussian profile (accounting for the doublet nature of the line), several events (SNe 2020auv, 2020onb, 2020wnt, 2024amf, 2024jwx) exhibit an additional narrow component which is in some cases blueshifted by $\approx2000$ km s$^{-1}$.  In addition, a few events exhibit two broad components, one located near the rest wavelength of [\ion{O}{1}] or slightly blueshifted and the other redshifted by $\approx4000$ km s$^{-1}$ (SNe 2018avk, 2020wnt, 2024jlc, 2024ahr).  As can be seen in Figure \ref{fig:OI}, this is apparent as a significant red excess making the overall line profiles for these events asymmetric.

Several events in our sample were observed at multiple epochs that captured the development of the emission line profile as the spectra transition to the nebular phase.  For example, in the early epochs for SNe 2020abjc, 2020znr, 2020qlb, and 2022le, we see photospheric absorption features on top of a larger broad profile with a peak flux to the red of the rest wavelength of [\ion{O}{1}], indicating the spectra are not fully nebular.  The later epochs for these events show a Gaussian emission profile developing and decreased shift; for example, the line is centered on the rest wavelength of [\ion{O}{1}] in the second epoch of SN\,2020qlb.  To further illustrate this timescale effect on the shape of the line, we plot the FWHM and central wavelengths derived from our line fits in Figure \ref{fig:OIvel} (including the narrow components if present).  Broad profiles with redshifted central wavelengths correspond to spectra taken at relatively early phases, whereas narrower profiles close to 6300 \AA\ are observed at later phases when the line has more fully developed.

\subsubsection{7300 \AA\ Feature}
\label{sec:7300}

In Figure \ref{fig:CaII} we show our fits to the 7300 \AA\ feature, often attributed to the [\ion{Ca}{2}] $\lambda$7300 doublet.  In Figure \ref{fig:CaIIvel} we show the corresponding FWHM and line centers for the 7300 \AA\ feature fits.  In general, the 7300 \AA\ feature shows greater complexity than the 6300 \AA\ feature, with many events exhibiting clear multi-component profiles.      

SNe 2020wnt and 2024dde show redshifted narrow components and a blueshifted broad base, whereas SNe 2020abjc, 2020znr, 2022le, 2024jlc, and 2024ahr show narrow components centered near the rest wavelength of [\ion{Ca}{2}] combined with a broad component centered to the red of [\ion{Ca}{2}].  In SN\,2020abjc this manifests as a clear shoulder to the red of a narrow Gaussian-like peak.  By the last epochs of SNe 2020abjc, 2020znr, and 2022le, these events are best fit by a single Gaussian centered close to [\ion{Ca}{2}], indicating a decreasing contribution from the red components with time.    

The red components in these objects suggests that emission from [\ion{O}{2}] $\lambda7320,7330$ is contributing to the 7300 \AA\ feature in many cases.  SNe 2022le and 2020abjc also show clear lines of [\ion{O}{3}] $\lambda4959,5007$ and [\ion{O}{3}] $\lambda4363$, supporting this conclusion (Figure \ref{fig:allspec}).  There is also a clear difference in width between the components centered near [\ion{Ca}{2}], which are narrow, and the red components, which are generally broad.  We further explore the presence of emission from ionized oxygen and the properties of the [\ion{O}{3}] lines in Section \ref{sec:ratios}.

Figure \ref{fig:CaIIvel} also shows that lines best fit by single components exhibit minimal shifts of $\sim$few$\times100$ km s$^{-1}$, though some are redshifted by up to $\approx1000$ km s$^{-1}$, possibly also an indication of emission from [\ion{O}{2}].  

Overall, the component line widths range from $\approx$1000 km s$^{-1}$ to $\approx$10,000 km s$^{-1}$, though the components centered near [\ion{Ca}{2}] are all less than $\approx7500$ km s$^{-1}$.  Among these, some are as low as $\approx700-2000$ km s$^{-1}$.        

\subsubsection{7774 \AA\ Feature}
\label{sec:7774}

In Figure \ref{fig:OI7774} we show our fits to the region containing the prominent oxygen recombination line \ion{O}{1} $\lambda7774$, which is detected in most events.  No obvious emission is seen in SN\,2018bym, SN\,2020aewh, SN\,2020auv, and SN\,2020tcw.  SN\,2020tcw exhibits significant host contamination while the other three have low S/N in this region of the spectrum.  We are therefore unable to rule out \ion{O}{1} $\lambda7774$ emission in these events.  The rest of the sample shows strong emission in this region, some with profiles centered near 7774 \AA\ (e.g., SN\,2018avk) and others with a prominent red shoulder (e.g., SN\,2020abjc) that sometimes also has an additional emission peak (e.g., SN\,2020znr).  Overall the profiles are complex, with most events exhibiting a red shoulder and/or signs of multiple components.

In Figure \ref{fig:OI7774vel} we show the FWHM and central wavelengths of the fitted components.  Profiles that are best fit by a single Gaussian have widths of $\approx2000-7000$ km s$^{-1}$ and are centered near 7774 \AA, though some show redshifts of $\approx500-1500$ km s$^{-1}$.  Profiles that are best fit by two or three Gaussian components generally have at least one component centered near 7774 \AA\ with widths of $\approx1000-5000$ km s$^{-1}$ along with a broad and/or narrow redshifted component with widths of $\approx1000-7000$ km s$^{-1}$. 

As noted in \citet{Nicholl2019}, there is a \ion{Mg}{2} doublet located at 7877 and 7896 \AA\ that was suggested as an explanation for the red shoulders of emission seen in about half of that sample.  Here we see clear evidence that \ion{Mg}{2} is indeed the source of the red emission.  In several events there are clear emission components centered near \ion{Mg}{2} which decrease with time, as most clearly demonstrated by SN\,2020abjc and SN\,2020znr.  Combined with the fact that some events show only a single component centered closer to 7774 \AA\ indicates that there is likely a range of possible \ion{Mg}{2} strengths.  Although it is unclear what is causing this variation, when \ion{Mg}{2} is present it appears to probe similar regions of the ejecta as \ion{O}{1} $\lambda7774$ given the similar line widths.   

\begin{figure*}
\centering
\includegraphics[scale=0.60]{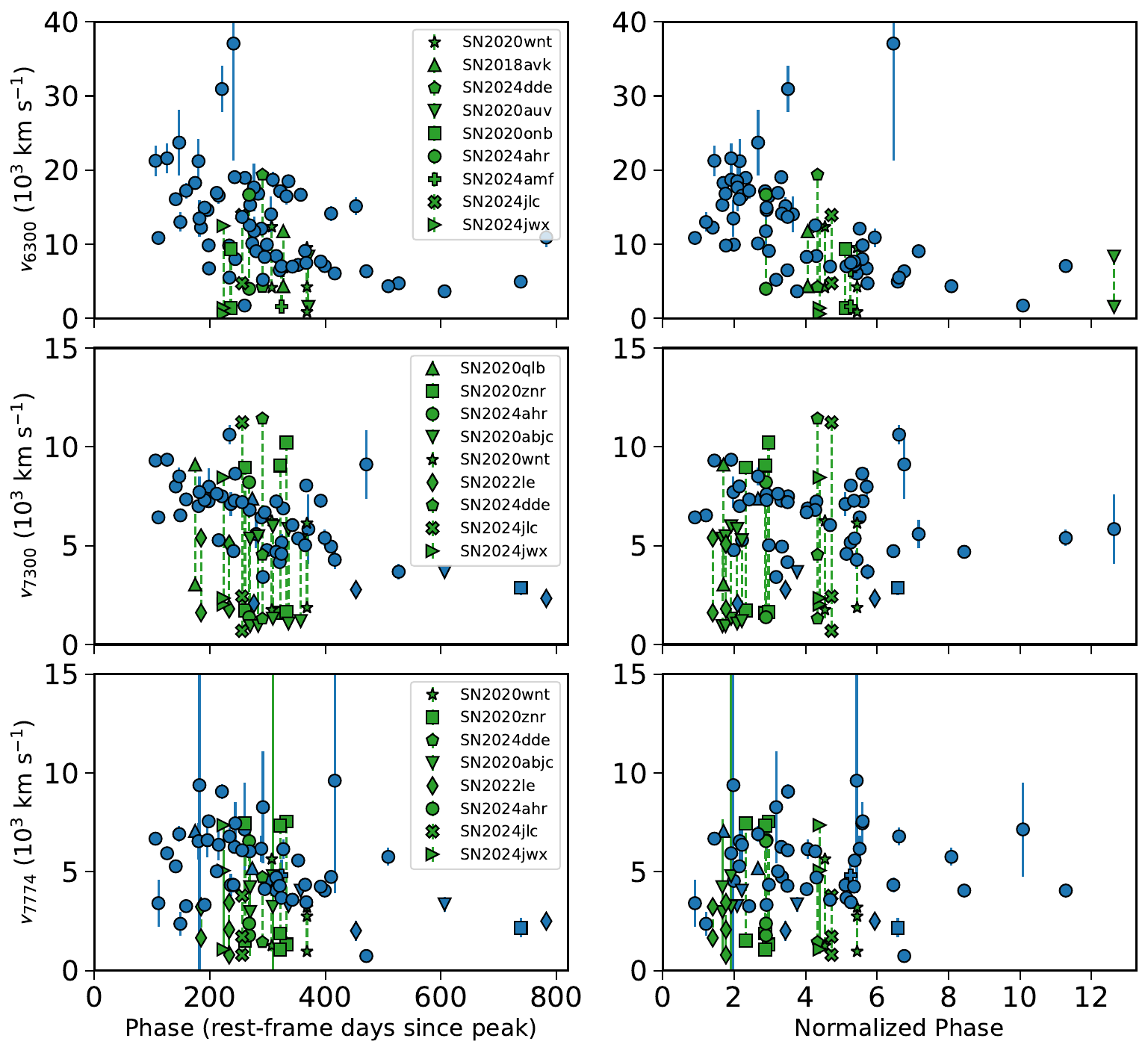}
\caption{Line width evolution from the fits to the profiles of the 6300 (top row), 7300 (middle row), and 7774 \AA\ (bottom row) features.  Profiles best fit with multiple components are shown in green with lines connecting the values for each component.  We show the velocities as a function of rest-frame days since peak (left column) and normalized phase (defined in the text; right column).  At early normalized phases of $t/t_{\rm d}\lesssim3$, the velocities of the 7300 and 7774 \AA\ features are much lower than the 6300 \AA\ feature.  The 6300 \AA\ feature eventually narrows, matching the 7300 and 7774 \AA\ velocities.}
\label{fig:velevol}
\end{figure*}

\subsubsection{Velocity Evolution}

In Figure \ref{fig:velevol} we show the velocity widths of the 6300, 7300, and 7774 \AA\ emission features from each spectrum as a function of rest-frame days since peak.  For those spectra with best-fit emission profiles containing more than one component, we show the widths of each component.  We also include in the figure the sample of 12 SLSNe from \citet{Nicholl2019}, which we analyzed in the same manner as our new spectra.  As this combined sample includes SLSNe with a wide range of light curve evolution timescales, we also plot the line velocities as a function of normalized phase, which is the rest-frame days since peak normalized by the characteristic decline timescale, $t_{\rm d}$.  We use the measured values of $t_{\rm d}$, defined as the time since peak for the luminosity to drop by a factor of $e$, from our recent light curve catalog presented in \citet{Gomez2024}.

The combined sample shows the same clear decreasing trend in the velocity evolution of [\ion{O}{1}] $\lambda6300$ as was seen in the \citet{Nicholl2019} sample.  The velocities are initially $\gtrsim10^{4}$ km s$^{-1}$ at normalized phases of $t/t_{\rm d}\lesssim3$ and smoothly decline to $\approx5000$ km s$^{-1}$ at $t/t_{\rm d}\gtrsim5$.  As noted in Section \ref{sec:6300}, the 6300 \AA\ features in some events were best fit with multiple components each with at least one relatively low velocity component ($\lesssim5000$ km s$^{-1}$).  Those components with the lowest velocities of $\approx1000-1500$ km s$^{-1}$ were detected in spectra obtained at late normalized phases, indicating such low-velocity material is only visible at these wavelengths at late times.

When the [\ion{O}{1}] velocities are still high at $t/t_{\rm d}\lesssim3$, the velocities of the 7300 and 7774 \AA\ features are much lower, ranging from $\approx1000-9000$ km s$^{-1}$.  As noted in Section \ref{sec:7300}, several events show signs that the 7300 \AA\ feature contains a contribution from [\ion{O}{2}].  In these cases both components are narrower than [\ion{O}{1}] at the same phase, with the component centered on [\ion{Ca}{2}] being substantially narrower at $1000-2000$ km s$^{-1}$.  These objects, along with SN\,2020wnt, also show such narrow components of \ion{O}{1} $\lambda7774$ and in some cases similarly narrow \ion{Mg}{2}.  Those events which are best fit with single components for the 7300 and 7774 \AA\ features also show substantially lower velocities than [\ion{O}{1}] at the same phase. 

Other than a mild decrease in the velocity of the 7300 \AA\ feature at $t/t_{\rm d}\gtrsim5$, there is no significant evolution of the 7300 and 7774 \AA\ features.  These lines are therefore probing low-velocity material at all times.  The early appearance of narrow [\ion{Ca}{2}]/[\ion{O}{2}], \ion{O}{1}, and \ion{Mg}{2} while [\ion{O}{1}] is broad indicates some of the low-velocity inner ejecta is becoming nebular and visible to the outside at timescales of only $t/t_{\rm d}\sim2$.  Eventually the velocity of the [\ion{O}{1}] emitting material reaches similar velocities as that emitting [\ion{Ca}{2}]/[\ion{O}{2}], \ion{O}{1}, and \ion{Mg}{2}. 

\subsection{Line Ratios and Signs of Ejecta Ionization}
\label{sec:ratios}

The [\ion{Ca}{2}] $\lambda$7300 and [\ion{O}{1}] $\lambda$6300 lines have been extensively studied in normal SNe Ic (e.g., \citealt{Filippenko1997,Matheson2001,Taubenberger2009,Milisavljevic2010,Fang2019,Fang2022,Prentice2022}) as they are often the strongest nebular lines due to the high abundance of oxygen and the efficiency of calcium as a coolant.  Theoretical models, including those generated with the goal of reproducing SLSNe, show that the ratio of these lines is sensitive to the core mass of the progenitor star \citep{FranssonChevalier1989,Jerkstrand2015,Jerkstrand2017,Dessart2021}.  The first sample of SLSN nebular spectra exhibited ratios that are on average consistent with more massive cores than for SNe Ic \citep{Nicholl2019}. 

However, the models also show that the level of ejecta ionization (predicted to be high in magnetar-powered explosions) can significantly affect both the strength of [\ion{O}{1}] $\lambda$6300 and the strength of the feature at 7300 \AA, which, in the case of high ionization, can be almost entirely dominated by [\ion{O}{2}] $\lambda$7320,7330 \citep{Jerkstrand2017,Dessart2019,Dessart2024,OmandJerkstrand2023}.  In these cases, the ratio does not correspond to [\ion{Ca}{2}]/[\ion{O}{1}].  These models show that a number of factors, including the explosion energy, deposition profile, and degree of clumping, influence the level of ionization and therefore the ratio of the 7300 \AA\ to the 6300 \AA\ features.  In addition, the phase of observation is another factor influencing the relative strength of these lines, a fact more relevant for SLSNe due to their overall slower evolution and larger diversity in evolution timescales compared to SNe Ic.  With these factors in mind, in this section we assess $L_{7300}/L_{6300}$ and other key ratios.  We assume that host reddening is negligible, a reasonable assumption given that photometric modeling and spectroscopic studies of SLSN host galaxies find little to no host extinction \citep{Lunnan2014,Perley2016,Schulze2018}.  

\begin{figure*}[t]
\centering
\includegraphics[scale=0.60]{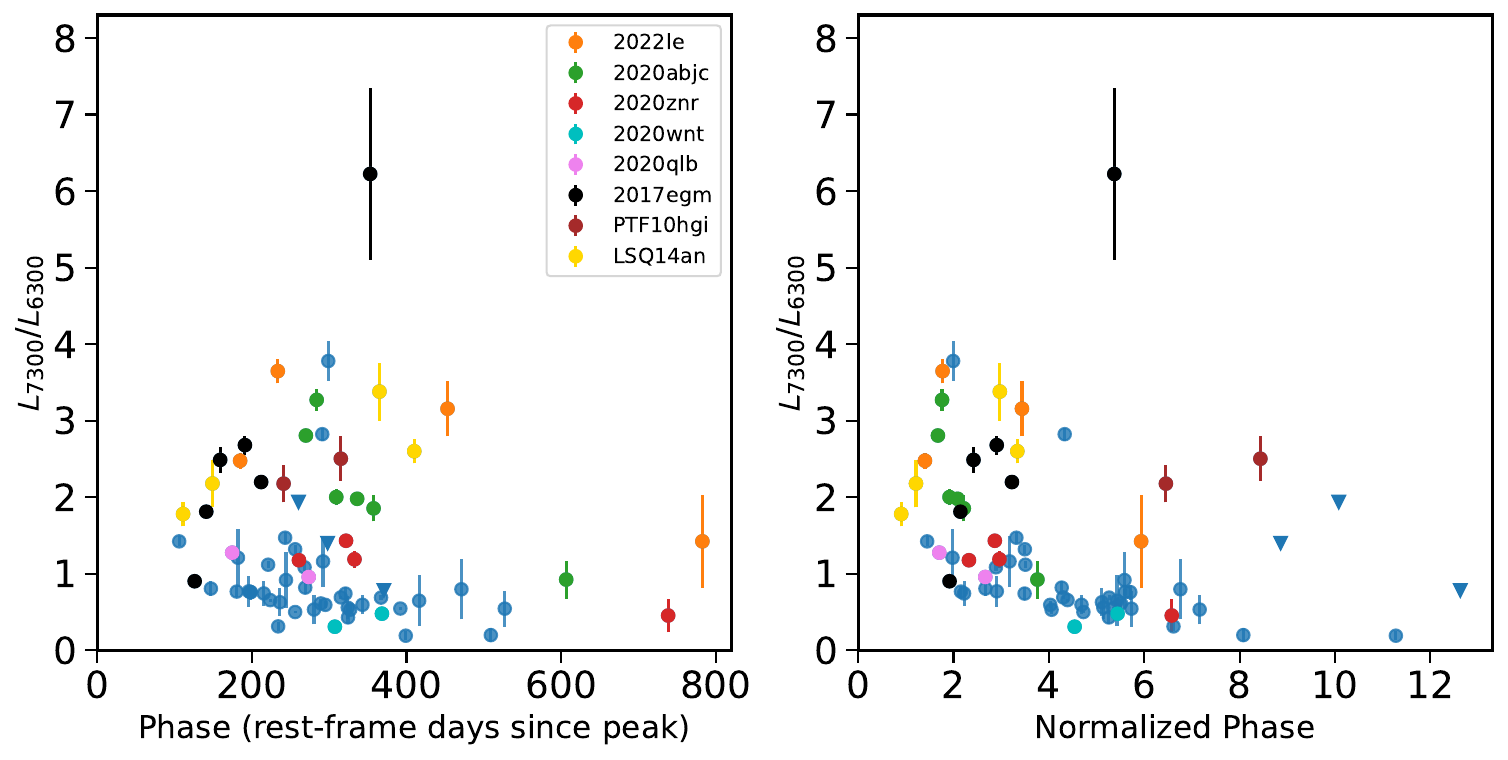}
\caption{Emission-line ratio $L_{7300}/L_{6300}$, which we show is a proxy for ejecta ionization (Section \ref{sec:ratio1}), as a function of rest-frame phase (left) and phase normalized by $t_{d}$ (right); selected events discussed in the text are labeled, while the rest of the sample are colored blue.  SLSNe show extreme diversity in this key ratio ($\approx 0.7-3.8$ at $t/t_{d}\lesssim4$), unlike normal SNe Ibc and Ic-BL, which nearly always have ratios below unity.  As discussed in Section \ref{sec:comps}, this is not simply due to a timescale effect, but rather reflects an intrinsic difference between SLSNe and their lower-luminosity counterparts.  $L_{7300}/L_{6300}$ shows an overall decreasing trend with time, with a few events exhibiting the opposite trend.}
\label{fig:ratiotime}
\end{figure*}

\begin{figure*}[!]
\centering
\includegraphics[scale=0.45]{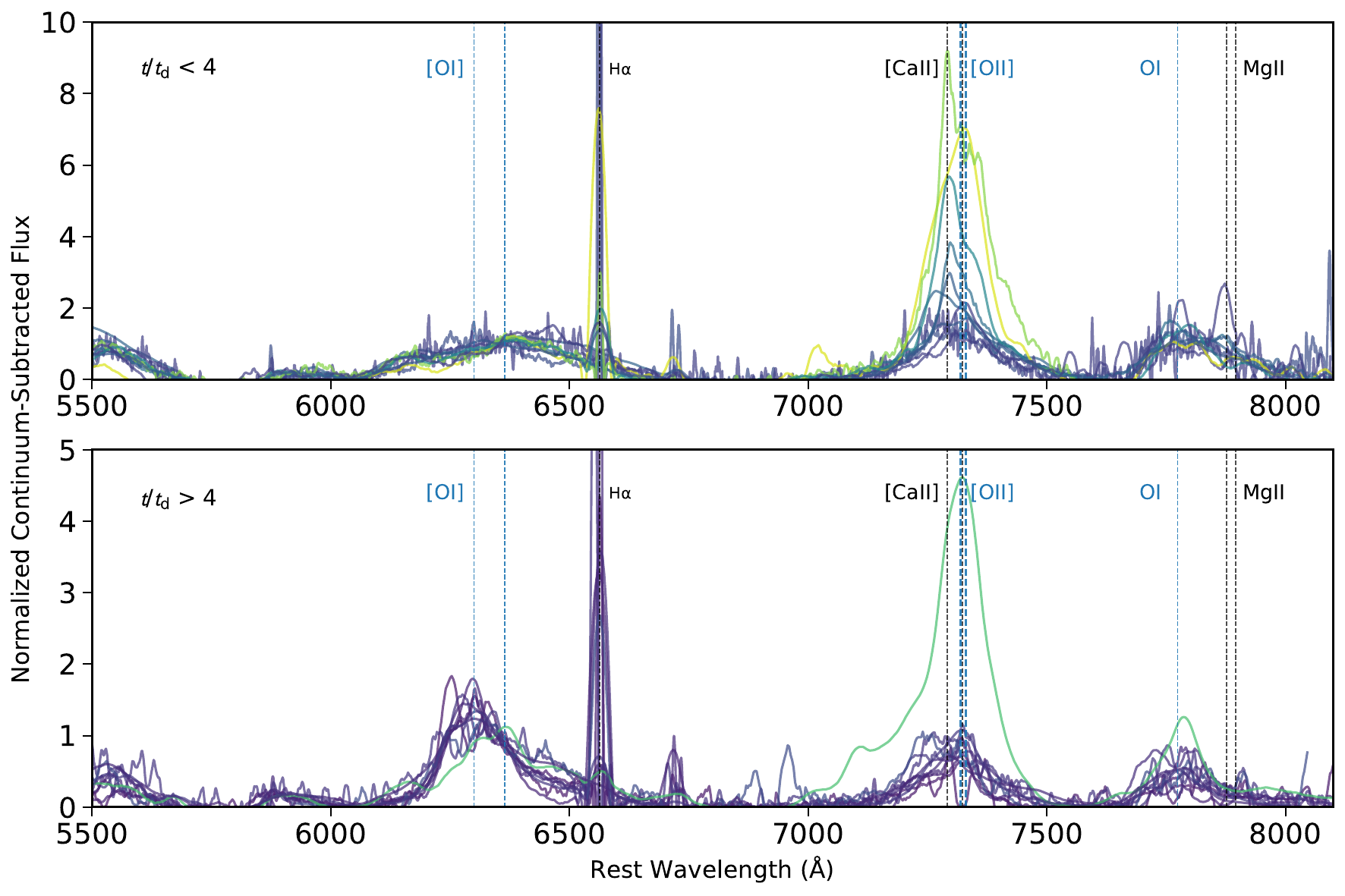}
\caption{Example spectra zoomed-in to the $5500-8000$ \AA\ region, divided into early ($t/t_{d} < 4$) and late ($t/t_{d} > 4$) normalized phases, demonstrating the significant variation in $L_{7300}/L_{6300}$ seen at early phases.}
\label{fig:specratiopanel}
\end{figure*}

\begin{figure*}[!]
\centering
\includegraphics[scale=0.5]{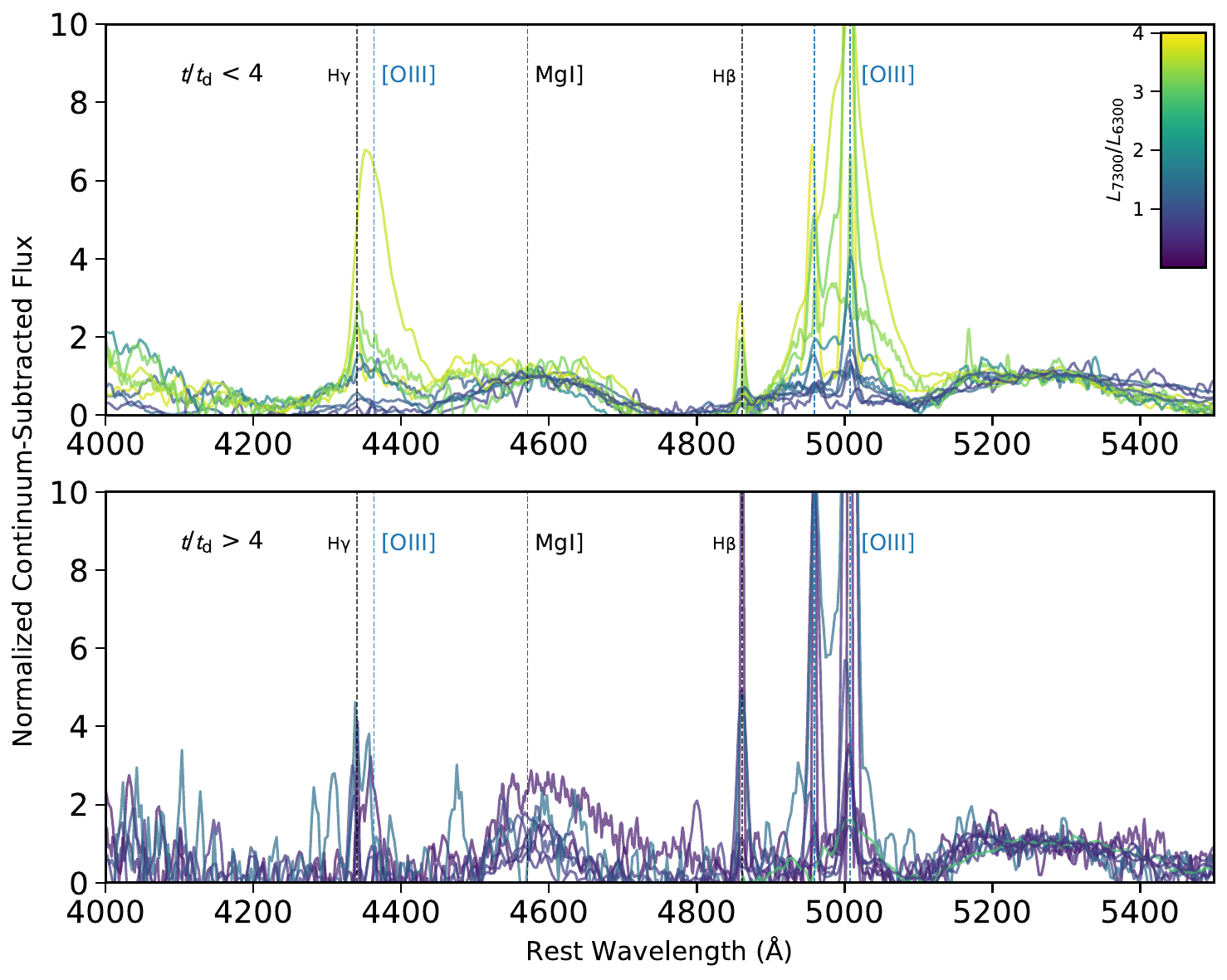}
\caption{Example spectra zoomed-in to the $4000-5500$ \AA\ region, divided into early ($t/t_{d} < 4$) and late ($t/t_{d} > 4$) normalized phases, demonstrating that the strength of [\ion{O}{3}] emission is correlated with $L_{7300}/L_{6300}$.  Events with a clear contribution of [\ion{O}{2}] to the 7300 \AA\ feature and therefore high $L_{7300}/L_{6300}$ also exhibit strong [\ion{O}{3}]$\lambda4363$ and [\ion{O}{3}]$\lambda\lambda4959,5007$.}
\label{fig:specOIII}
\end{figure*}

\subsubsection{$L_{7300}/L_{6300}$ and the Presence of [\ion{O}{2}] and [\ion{O}{3}] Lines}
\label{sec:ratio1}

As evident in Figure \ref{fig:allspec}, the ratio $L_{7300}/L_{6300}$ spans a broad range.  To understand this diversity we calculate this ratio from each spectrum by integrating the best-fit line profiles (\S\ref{sec:6300} and \S\ref{sec:7300}, and Figures \ref{fig:OI} and \ref{fig:CaII}).  In Figure~~\ref{fig:ratiotime} we plot $L_{7300}/L_{6300}$ both as a function of phase and normalized phase.  Included in the figure are the ratios for the previous sample of \citet{Nicholl2019}, which we measured using our same line-fitting procedure.  The resulting values span $\approx0.2$ to $\approx 6.2$, with a mean of $\approx 1.4$, with clear evidence for temporal evolution among the sample as a whole, but with diverse trends as evident from individual objects with multiple spectra.  Our sample provides measurements at the latest timescales to date of this ratio, with the latest absolute phase measurement from our spectrum of SN\,2020znr at $+729$ rest-frame days post-peak and the latest normalized phase measurement of SN\,2020auv at $t/t_{\rm d} \approx 12.6$.  Overall, there is a trend of decreasing $L_{7300}/L_{6300}$ with time (e.g., SN\,2020abjc, SN\,2020znr, and SN\,2020qlb, as well as previously in SN\,2015bn, \citealt{Nicholl2019}).  Given the broad range of evolutionary timescales, this decreasing trend is most clearly evident in the normalized phase.  The normalized phase also reveals that the ratio exhibits greater diversity at earlier times.  At $t/t_{\rm d} < 4$ $L_{7300}/L_{6300}$ spans $\approx0.7-3.8$ with a mean of $\approx2$, whereas at $t/t_{\rm d} > 4$ the ratio is typically below unity with continued decreasing to very late normalized phases (except for several outliers; SN\,2024dde, SN\,2017egm, and PTF10hgi).

To understand what is driving the wide range of line ratios, we plot example spectra in Figure \ref{fig:specratiopanel}.  The events with high $L_{7300}/L_{6300}$ are those for which we identify a contribution from [\ion{O}{2}] to the 7300 \AA\ feature (events with an additional red component or a single component centered  redward of [\ion{Ca}{2}]).  This is strong evidence that the high ratios are due to the presence of ionized oxygen, rather than unusually strong [\ion{Ca}{2}].  

The presence of ionized oxygen is further supported by the detection of [\ion{O}{3}]$\lambda4364$ and [\ion{O}{3}]$\lambda\lambda4959,5007$ in the events with the highest $L_{7300}/L_{6300}$ ratios (SN\,2020abjc, SN\,2020znr, SN\,2022le, SN\,2018fd, LSQ14an); PS1-14bj \citep{Lunnan2016} and LSQ14an \citep{Jerkstrand2017,inserra_complexity_2017} were previously identified to have these lines.  In Figure \ref{fig:specOIII} we show selected spectra zoomed-in to the [\ion{O}{3}] features.  The velocities of the [\ion{O}{3}] lines span $\approx3500-6000$ km s$^{-1}$ and are therefore similar or slightly lower than the velocities of the [\ion{O}{2}] lines and slightly larger than the velocities of the \ion{O}{1}$\lambda7774$ lines.  SN\,2020abjc and SN\,2022le additionally exhibit blueshifted narrow components with widths of $\approx 750$ km s$^{-1}$ and blueshifts of $\approx1250$ km s$^{-1}$.

As noted above, several events deviate significantly from the trend of decreasing $L_{7300}/L_{6300}$ with phase, instead showing an increasing or constant ratio.  LSQ14an and SN\,2017egm exhibit particularly extended phases of increasing $L_{7300}/L_{6300}$, although due to LSQ14an's slow evolution, all of its spectra are at $t/t_{\rm d} < 4$.  This results in elevated ratios of $L_{7300}/L_{6300} \gtrsim2$ at $t/t_{\rm d} \approx 3$, whereas most events have $L_{7300}/L_{6300} \approx1$ by this normalized phase.  In its last spectrum, SN\,2017egm exhibits a further increase to the highest observed value of any SLSN.  PTF10hgi exhibits a fairly constant elevated ratio of $L_{7300}/L_{6300} \approx 2.5$ at $t/t_{\rm d} \approx  6.5-8.5$.  SN\,2024dde shows a high ratio of $L_{7300}/L_{6300} \approx 2.8$ at $t/t_{\rm d}\approx 4.3$.  We discuss the implications of these outliers in \S\ref{sec:csm}.

\subsubsection{$L_{7774}/L_{6300}$}

In Figure \ref{fig:otherratios} we show the ratio $L_{7774}/L_{6300}$ as a function of phase and normalized phase.  As discussed in \S\ref{sec:7774}, many events exhibit complex emission profiles in the 7774 \AA\ region with contributions from \ion{O}{1} and \ion{Mg}{2}.  In addition, the relative contribution of \ion{Mg}{2} can vary significantly.  To be consistent with previous work and to avoid systematic uncertainties associated with deblending the individual \ion{O}{1} and \ion{Mg}{2} contributions, we calculate $L_{7774}$ by integrating the full profile, regardless of the presence or absence of strong emission on the red side of \ion{O}{1}.  

This ratio also exhibits considerable diversity, spanning a  range of $\approx0.15-1.1$, with a mean of $\approx 0.5$.  The $L_{7774}/L_{6300}$ ratio generally shows a decline to later normalized phases, although some events maintain high ratios.  At least some of this diversity is likely due to the variable contribution of \ion{Mg}{2}.  As seen in SN\,2020znr and SN\,2020abjc, the relative contribution of \ion{Mg}{2} decreases with time, which may explain the slight decrease in $L_{7774}/L_{6300}$.  Therefore, the ratio at later phases is likely a better probe of the true \ion{O}{1} $\lambda7774$/[\ion{O}{1}] $\lambda6300$ ratio.  The clear detection of \ion{O}{1} $\lambda7774$ emission to late phases is a hallmark feature of the nebular spectra of SLSNe compared to SNe Ic, which we discuss further in \S\ref{sec:comps}.  

\subsubsection{$L_{7300}/L_{\rm CaII\,NIR}$}

The ratio of the two dominant \ion{Ca}{2} features, [\ion{Ca}{2}] $\lambda7300$ and the \ion{Ca}{2} NIR triplet, can be used to constrain the electron density.  Events with coverage of the NIR triplet, generally show strong emission (Figure~\ref{fig:allspec}).  Due to the complex nature of this spectral feature, with the three lines often blended and with varying ratios, we do not fit the profile as with the other lines.  Instead we measure $L_{\rm CaII\,NIR}$ by directly integrating the full blended profile. We estimate the continuum contribution by fitting a line to regions blueward and redward of the emission feature.  We place a lower limit on $L_{\rm CaII\,NIR}$ in cases where only part of the emission profile is detected due to limited wavelength coverage.

\begin{figure*}[t]
\centering
\includegraphics[scale=0.52]{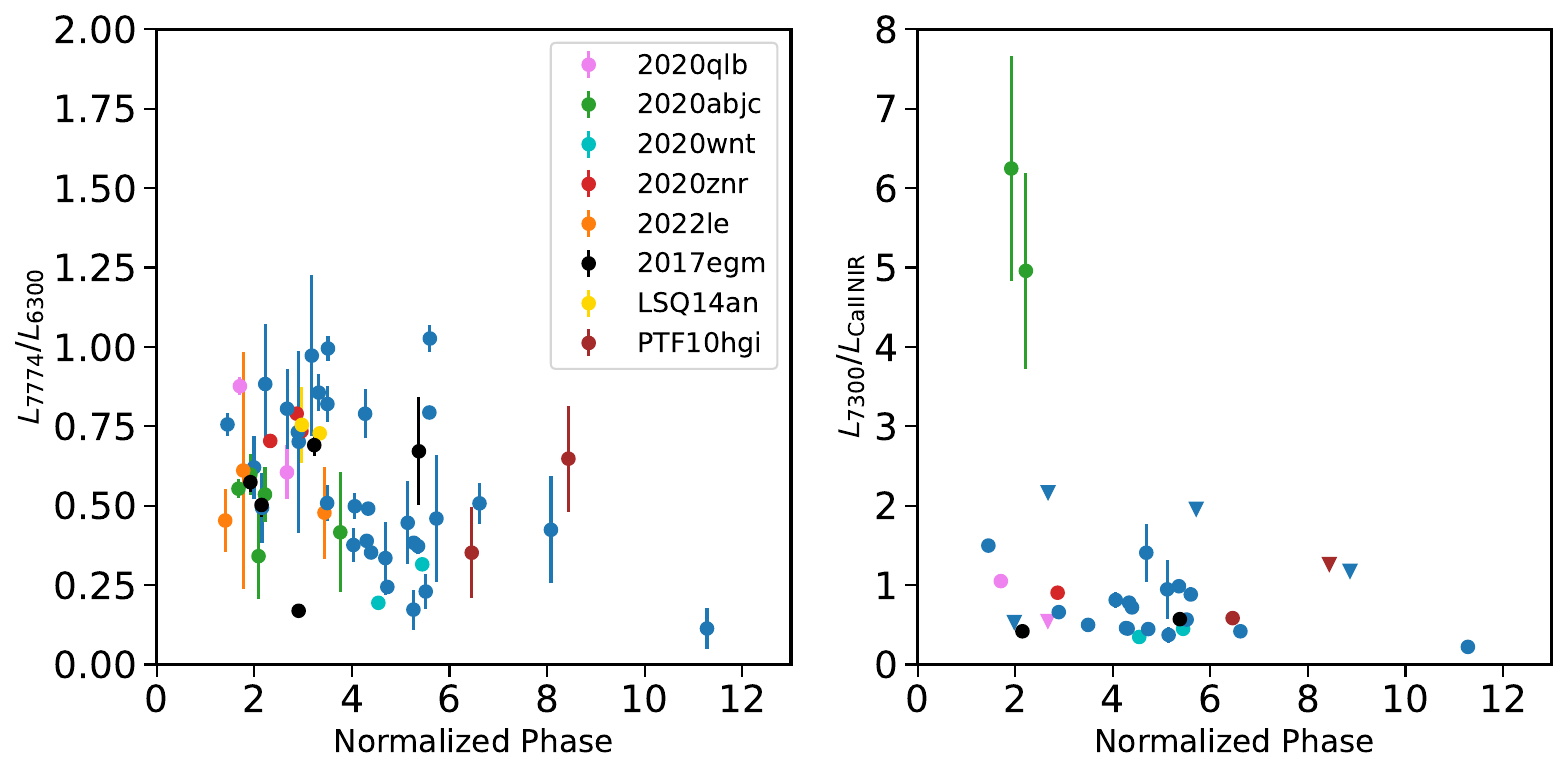}
\caption{Emission line ratios $L_{7774}/L_{6300}$ (left) and $L_{7300}/L_{\rm CaII\,NIR}$ (right) as a function of normalized phase.  $L_{7774}/L_{6300}$ exhibits a slight decreasing trend with time, possibly due to a decreasing contribution from \ion{Mg}{2}.  Even at late phases, SLSNe maintain strong \ion{O}{1} $\lambda7774$. SLSNe exhibit low $L_{7300}/L_{\rm CaII\,NIR}$ throughout their evolution.}
\label{fig:otherratios}
\end{figure*}

In Figure~\ref{fig:otherratios} we plot $L_{7300}/L_{\rm CaII\,NIR}$ as a function of normalized phase, and find that the ratio is generally $\lesssim 1$, with a mean value of $\approx 0.6$.  The ratio exhibits little evolution with time, although the event with $L_{7300}/L_{\rm CaII\,NIR}\approx0.2$ at a late normalized phase of $t/t_{\rm d}\approx11$ suggests the ratio may eventually decrease.  As noted in \S\ref{sec:7300} and \S\ref{sec:ratio1}, [\ion{O}{2}] is clearly present and blended with [\ion{Ca}{2}] in several events.  In some cases, [\ion{O}{2}] is likely dominating the flux of the 7300 \AA\ feature.  For those cases (e.g., SN\,2020abjc, SN\,2020znr, SN\,2017egm, PTF10hgi), the true [\ion{Ca}{2}] $\lambda7300$/\ion{Ca}{2} NIR ratio is likely lower than the measured ratio $L_{7300}/L_{\rm CaII\,NIR}$, although other lines can also contaminate the \ion{Ca}{2} NIR triplet.  The low values of $L_{7300}/L_{\rm CaII\,NIR}$ in our sample imply electron densities of $n_e\gtrsim10^8$ cm$^{-3}$ \citep{Jerkstrand2017}, which we discuss further in \S\ref{sec:clumping}.

\begin{figure*}[t]
\centering
\includegraphics[scale=0.55]{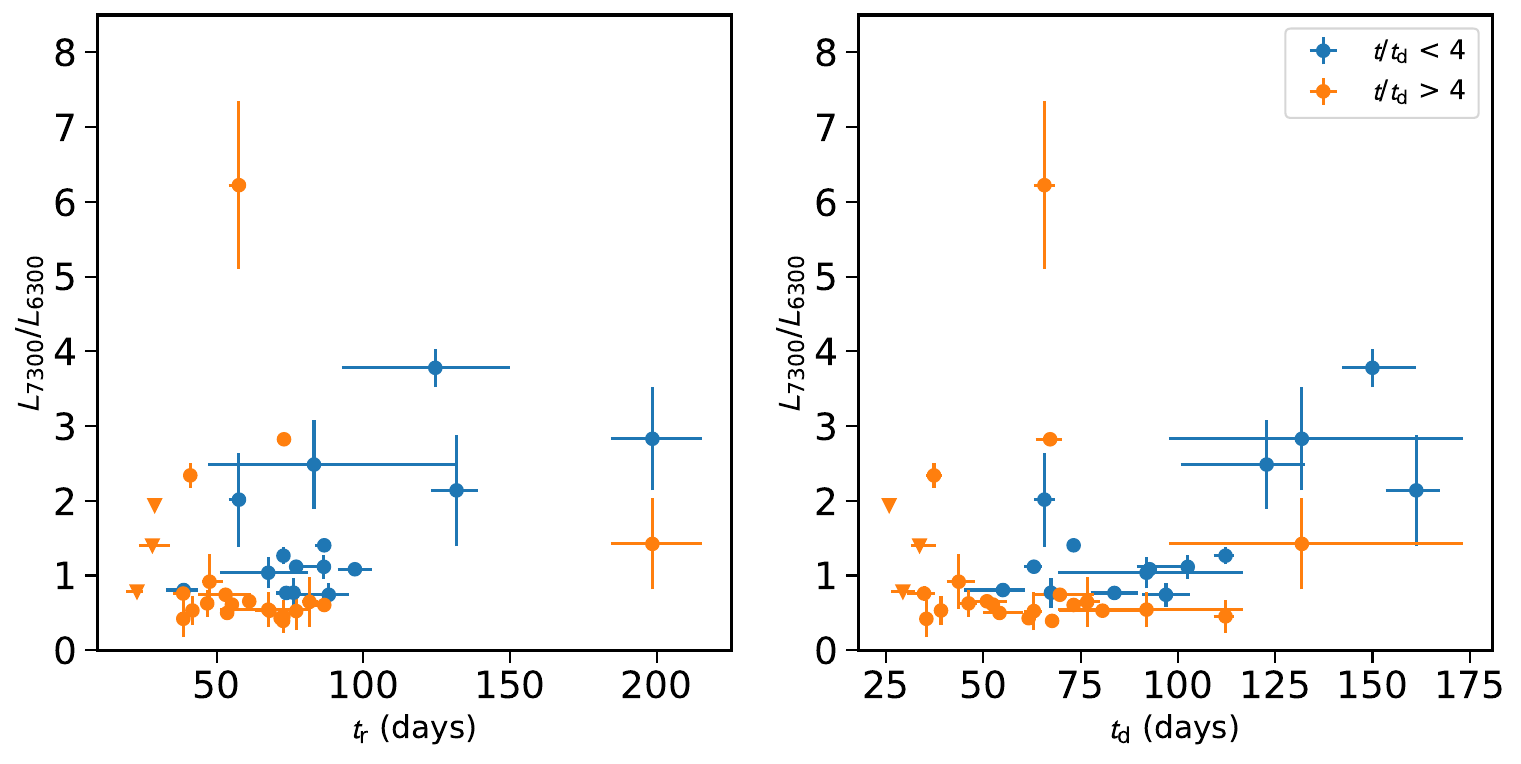}
\caption{Emission line ratio $L_{7300}/L_{6300}$ as a function of the rise (left) and decline (right) times.  The timescales are defined in the text.  To isolate the impact of spectral evolution, we label the sample based on normalized phase (blue for $t/t_{\rm d}<4$, orange for $t/t_{\rm d}>4$).  There is a clear correlation between $L_{7300}/L_{6300}$ and the two timescales, driven primarily by the sample at $t/t_{\rm d}<4$.  Among these, SLSNe with longer timescales have higher ratios.  In other words, the large spread in ratios is correlated with light curve timescale, the first robust correlation between light curve and nebular spectral properties identified in SLSNe.}
\label{fig:ratio_vs_risedec}
\end{figure*}

\section{Comparison with Light Curve and Magnetar Properties}
\label{sec:LCmag_comp}

Our large sample of nebular-phase spectra of SLSNe enables a comparison of their late-time spectral properties with their light curve properties (timescales, peak luminosities, radiated energies) and inferred engine and ejecta parameters (initial spin period, $P$; magnetic field strength, $B$; ejecta mass, $M_{\rm ej}$; kinetic energy, $E_{\rm K}$).  We use the properties inferred from our light curve modeling work in \citet{Gomez2024}.  

\subsection{Correlations Between $L_{7300}/L_{6300}$ and Light Curve Timescales}

In Figure~\ref{fig:ratio_vs_risedec} we show the line ratio $L_{7300}/L_{6300}$ as a function of the light curve rise and decline timescales.   For $t_{\rm r}$ we use the time from explosion, as inferred from {\tt mosfit} modeling, to $r$-band peak, while $t_d$ is the same $e$-folding time used throughout the paper.  Due to the significant temporal evolution in the line ratios, we first divide the spectra into two bins with phases $t/t_{d} < 4$ and $t/t_{d} > 4$.  For events with more than one spectrum in a given bin, we calculate the mean ratio.  

As evident in Figure~\ref{fig:ratio_vs_risedec}, there is a correlation between $L_{7300}/L_{6300}$ and both $t_{\rm r}$ and $t_{\rm d}$, with a Pearson correlation coefficient of $r \approx 0.34$ ($p \approx 0.03$) for $t_{\rm r}$ and $r \approx 0.39$ ($p \approx 0.01$) for $t_{\rm d}$.  Events with slower evolving light curves exhibit higher $L_{7300}/L_{6300}$.  The slowest events, with $t_{\rm r} \gtrsim 100$ days and $t_{\rm d} \gtrsim 125$, have $L_{7300}/L_{6300} \gtrsim 2$.  SLSNe with the most rapid timescales of $t_{\rm r} \lesssim 50$ days and $t_{\rm d} \lesssim 50$ have $L_{7300}/L_{6300}\lesssim 1$ (with PTF10hgi as a notable exception). Due to their rapid evolution, these events have spectra at later normalized phases of $t/t_{\rm d} > 4$ when even events with initially high $L_{7300}/L_{6300}$ have evolved to lower ratios.  Indeed, there is no correlation when considering only the spectra taken at late phases of $t/t_{\rm d} > 4$ ($r \approx 0.1$, $p \approx 0.6$).  The overall correlation is therefore driven by events with spectra at $t/t_{\rm d} < 4$ ($r \approx 0.63$, $p \approx 0.01$ for $t_{\rm r}$ and $r \approx 0.71$, $p \approx 0.003$ for $t_{\rm d}$).  Here it is evident that the spread is correlated with light curve timescale.  This suggests a fundamental difference between the physical conditions of the ejecta in faster and slower SLSNe.  We note that this is the first evidence that there exist early light curve properties of SLSNe (namely, $t_{\rm r}$) that predict the late-time spectral behavior.  

We do not find any correlations between the light curve timescales and the $L_{7774}/L_{6300}$ or $L_{7300}/L_{\rm CaII\,NIR}$ line ratios.  There are also no correlations between any of the line ratios and the peak luminosity or radiated energy.

\begin{figure*}
\centering
\includegraphics[scale=0.55]{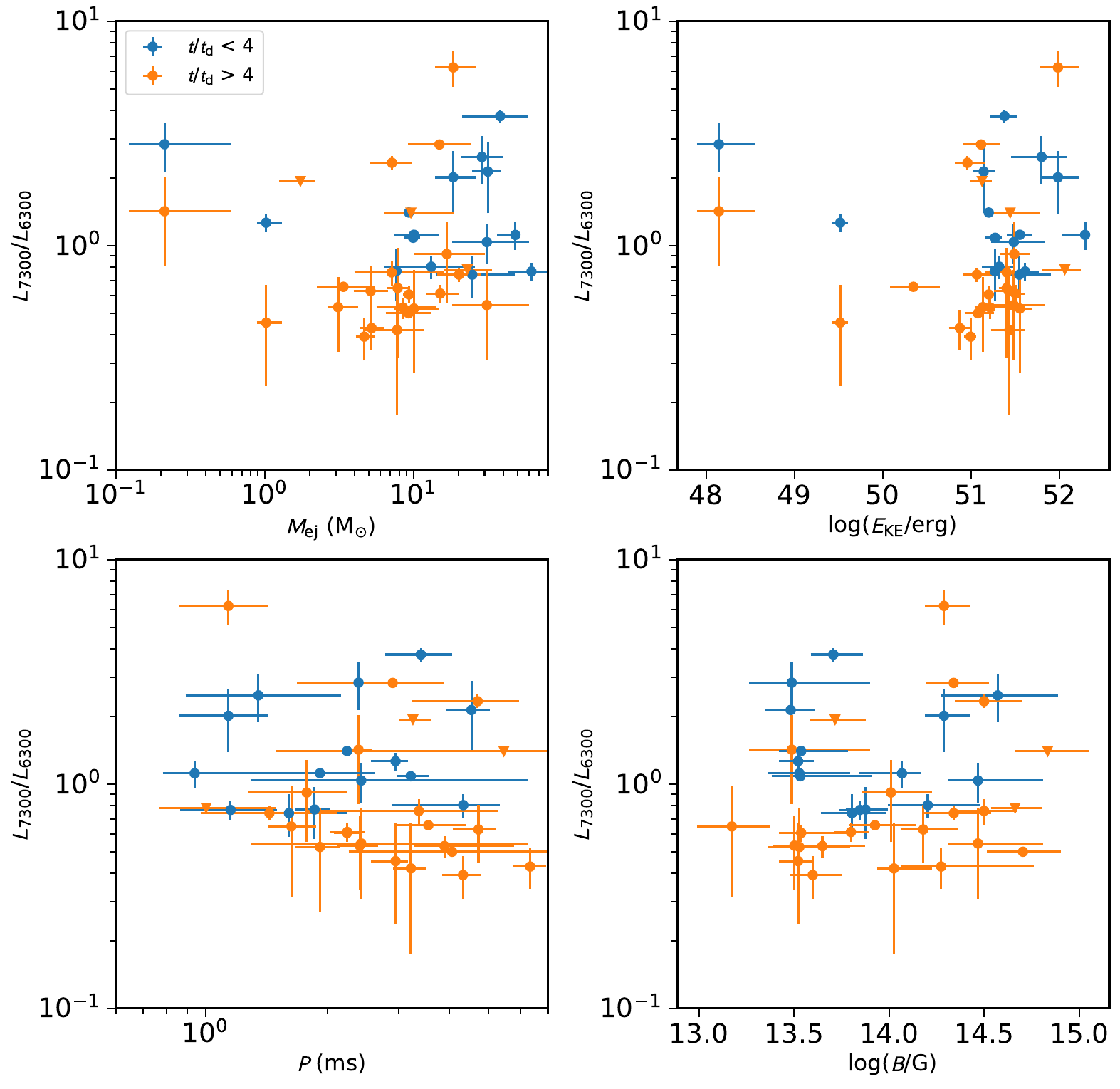}
\caption{Emission line ratio $L_{7300}/L_{6300}$ as a function of ejecta mass (top left), kinetic energy (top right), initial spin period (bottom left), and magnetic field strength (bottom right).  As in Figure \ref{fig:ratio_vs_risedec} we label the events in two normalized phase bins.  There are no statistically significant correlations with any of these parameters individually; we assess for correlations with magnetar spin-down timescale and instantaneous magnetar energy input (functions of $P$ and $B$) in Figures \ref{fig:tmag} and \ref{fig:magpower}.}
\label{fig:LCparams}
\end{figure*}

\begin{figure*}
\centering
\includegraphics[scale=0.55]{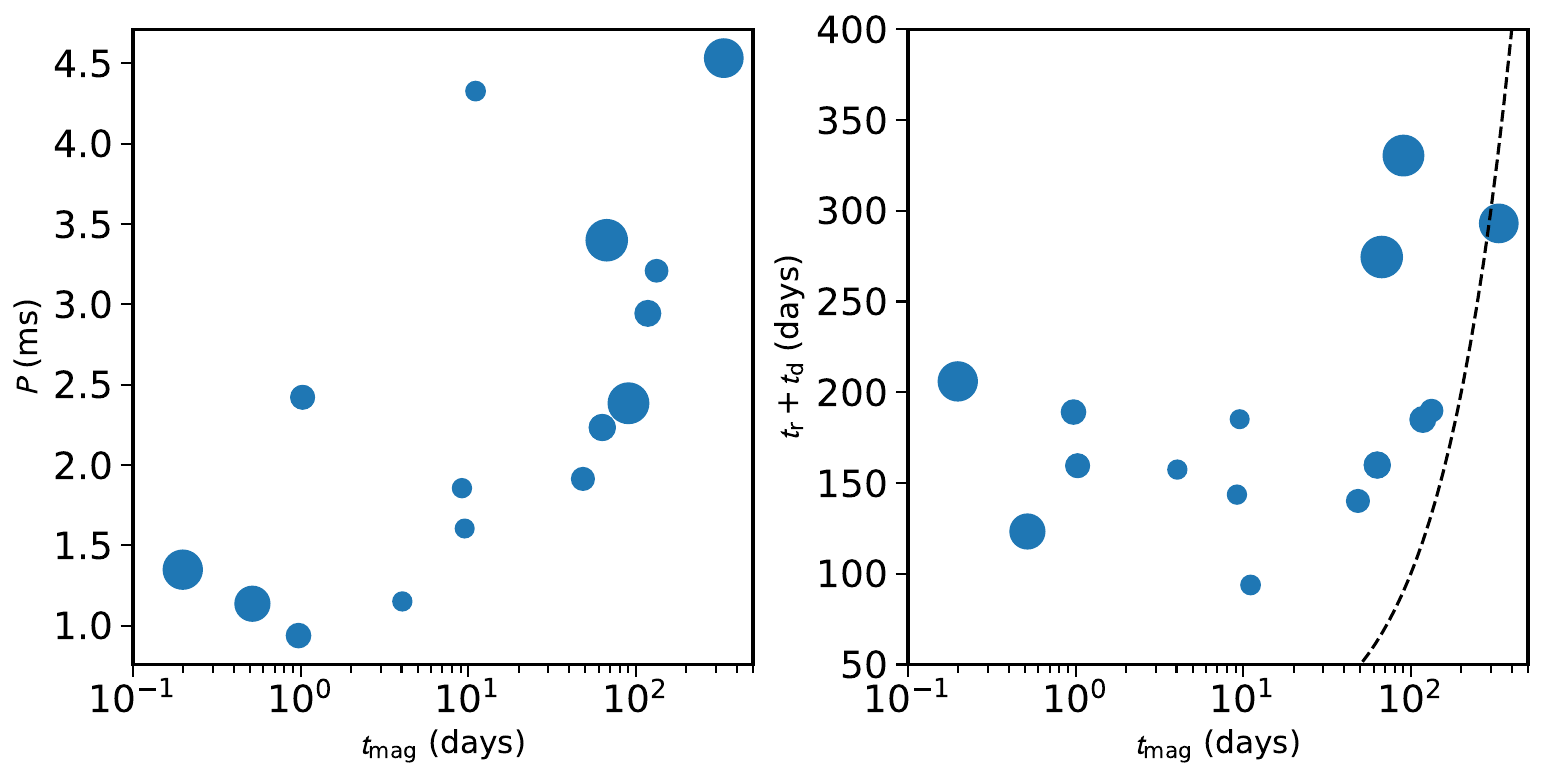}
\caption{Initial spin period ($P$; left) and the sum of the rise and decline timescale ($t_{\rm r}+t_{\rm d}$; right) as a function of spin-down timescale ($t_{\rm mag}$).  Only SLSNe with spectra at $t/t_{\rm d}\lesssim4$ are shown, the phase when there is a large spread in the ratio (Figure \ref{fig:ratiotime}).  The size of the points correspond to the ratio $L_{7300}/L_{6300}$.  The dashed line in the right panel corresponds to $t_{\rm r}+t_{\rm d}$ = $t_{\rm mag}$.  While there is no correlation between $L_{7300}/L_{6300}$ and $t_{\rm mag}$, high ratio events tend to have either large $t_{\rm mag}$ or low $P$.  Events with large $t_{\rm mag}$ are generally those for which $t_{\rm mag}$ is a better match to the diffusion time (using $t_{\rm r}+t_{\rm d}$ as a proxy).}
\label{fig:tmag}
\end{figure*}

\begin{figure*}
\centering
\includegraphics[scale=0.55]{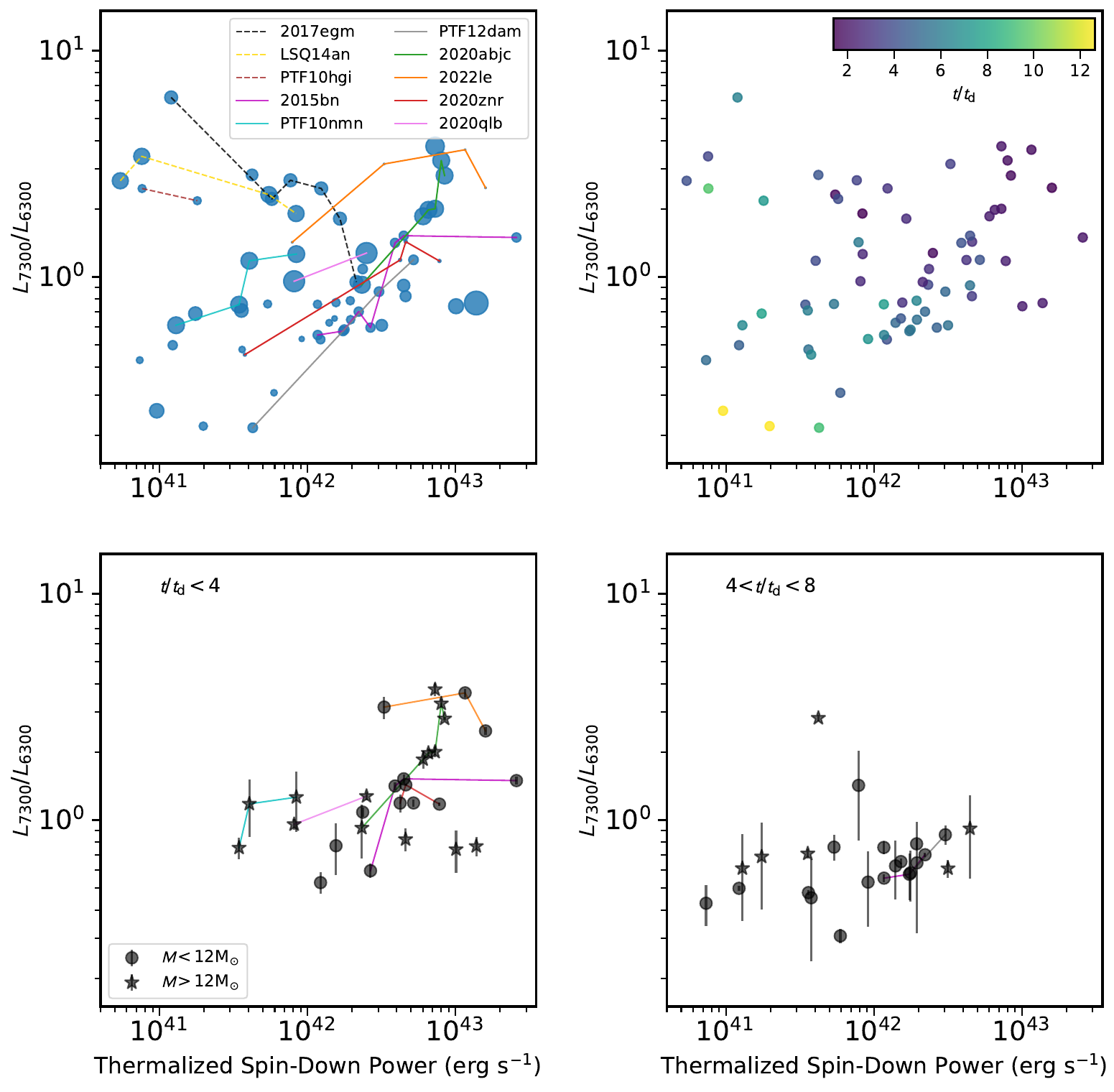}
\caption{$L_{7300}/L_{6300}$ as a function of the thermalized magnetar spin-down power (i.e., the instantaneous spin-down power absorbed by the ejecta) at the time of each spectrum.  We show the full sample with point sizes proportional to $M_{\rm ej}$ (top left) and colored by normalized phase (top right), as well as two subsets binned by normalized phase (bottom row). Selected events with multiple epochs of spectra are connected by lines, demonstrating the dominant trend of decreasing ratio with decreasing spin-down power.  Events in this population that have higher ratios, and therefore higher ionization, are on average powered by magnetars with higher spin-down power, a correlation that is most significant at $M_{\rm ej}\lesssim12$ M$_{\odot}$.  Events with increasing ratios (shown in the top panels only) are anti-correlated with the spin-down power, suggesting the ionization in these events is due to another effect.}
\label{fig:magpower}
\end{figure*}

\subsection{Comparison with Ejecta and Engine Parameters}

In Figure~\ref{fig:LCparams} we plot $L_{7300}/L_{6300}$ as a function of $M_{\rm ej}$, $E_{\rm K}$, $P$, and $B$ inferred from the light curve.  We find no statistically significant correlations between this ratio and the physical properties.  While a correlation with $M_{\rm ej}$ might be expected given the correlation with timescale, the engine spin-down timescale ($t_{\rm mag}$; equation 4 in \citealt{Nicholl2017}), governed by $P$ and $B$, also affects the light curve timescale; even events with relatively low ejecta masses can exhibit long-duration light curves if the magnetar's energy is deposited slowly. In addition, we do not find any correlations between these physical parameters and the $L_{7774}/L_{6300}$ or $L_{7300}/L_{\rm CaII\,NIR}$ line ratios.        

While there is no clear correlation with $P$, which sets the total energy available from the magnetar, or with the engine timescale, these factors may influence which events are able to reach high levels of ejecta ionization and therefore high ratios of $L_{7300}/L_{6300}$.  In Figure \ref{fig:tmag} we show the ratio, for spectra taken at $t/t_{\rm d} < 4$, relative to $t_{\rm mag}$ versus $P$ and $t_{\rm r} + t_{\rm d}$.  Events with long engine timescales of $t_{\rm mag} \gtrsim 50$ days have higher ratios ($L_{7300}/L_{6300} \gtrsim 1.5$) than events with $t_{\rm mag} \sim 1-50$ days.  Interestingly, events with $t_{\rm mag} \lesssim 1$ day also have higher ratios than events in the intermediate range.  Such events also have rapid initial spins and therefore a large reservoir of engine energy.  Magnetars with long engine timescales deposit more energy at later times than those with short timescales, with the overall energy scale set by the initial spin.  Therefore, the high ratios for events with long engine timescales and rapid initial spins may be due to larger instantaneous magnetar energy input.

\subsection{Correlation with the Thermalized Spin-Down Power}

In Figure~\ref{fig:magpower} we plot the thermalized instantaneous spin-down power at the time each spectrum was taken as a function of $L_{7300}/L_{6300}$.  We account for leakage of high-energy photons via the leakage term $1-e^{-At^{-2}}$, where $A$ is the leakage parameter \citep{Wang2015},
$A = 3\kappa_{\gamma}M_{\rm ej}/{4\pi v_{\rm ej}^2}$, where $\kappa_{\gamma}$ and $v_{\rm ej}$ are the gamma-ray opacity and the ejecta velocity, respectively.  The leakage term determines the fraction of the magnetar's energy that is absorbed by the ejecta.  

The general population shows a clear correlation between $L_{7300}/L_{6300}$ and the thermalized spin-down power.  For example, SN\,2015bn, SN\,2020znr, SN\,2020abjc, and SN\,2022le exhibit high ratios at early phases when the magnetar's energy input is high and then evolve to lower ratios as the energy input decreases.  This is not surprising given that both quantities decrease with time; in \S\ref{sec:ratios} we found that the SLSN population generally exhibits decreasing $L_{7300}/L_{6300}$ with time (SN\,2017egm, PTF10hgi, and LSQ14an being notable exceptions with increasing ionization, which we discuss in \S\ref{sec:csm} is likely due to an additional source of ejecta ionization beyond magnetar input alone).

To determine whether greater thermalized magnetar energy input leads to higher ejecta ionization across the sample, we analyze events in two bins of normalized phase.  For events with spectra at $t/t_{\rm d}<4$, we find a marginal correlation between the magnetar power and $L_{7300}/L_{6300}$ with a correlation coefficient of $r\approx0.39$ and $p\approx0.17$.  Notably, all of the sources with $L_{7300}/L_{6300}>2$ have spin-down power $\gtrsim7\times10^{42}$ erg s$^{-1}$, whereas the average spin-down power for events with $L_{7300}/L_{6300}\approx1$ is $\sim2\times10^{42}$ erg s$^{-1}$. 

In Figure \ref{fig:magpower}, we also search for any trends associated with $M_{\rm ej}$, which affects the spin-down power available per unit mass.  At the same phases of $t/t_{\rm d}<4$ but only considering events with $M_{\rm ej}<12$ M$_{\odot}$ (the median in this phase bin), we find a correlation coefficient of $r\approx0.97$ with $p\approx10^{-3}$, indicating a highly significant connection between the spin-down power and ejecta ionization for the lower mass half of the sample.  This is the first evidence of a direct link between the magnetar properties, as derived from the early light curves, and late-time spectral properties.  We note that this is robust to the precise choice of mass cut-off.  In other words, most sources support this correlation; it is only a few of the most massive events with $M_{\rm ej}\gtrsim20$ M$_{\odot}$ that deviate (namely the two events with $L_{7300}/L_{6300} \approx 0.8$ but high spin-down power of $\approx10^{43}$ erg s$^{-1}$).  This is perhaps not surprising as more massive events have on average lower spin-down power {\it per unit mass}; therefore, the engine's influence on ejecta ionization is reduced.  We also note that this correlation holds at later phases as well ($r\approx0.55$, $p\approx0.04$).  Importantly, we also test whether the integrated spin-down energy released up to a given time has any effect; we found no correlation with the $L_{7300}/L_{6300}$ ratio, indicating it is the instantaneous magnetar power that matters.

\section{Discussion} 
\label{sec:disc}

\subsection{Magnetar-Driven Ejecta Ionization}
\label{sec:ionization}

It has long been expected that a newly-formed pulsar wind nebula in a core-collapse SN can photoionize the surrounding SN ejecta \citep{ChevalierFransson1992}. More recent simulations specific to the magnetar engines relevant to SLSNe indicate they may drive ionization fronts that break out after completely ionizing the SN ejecta \citep{Metzger2014}, most likely engines with slowly decaying or constant ionizing luminosities \citep{Margalit2018}.  Given that oxygen is the most abundant element in the ejecta, strong lines of ionized oxygen are expected in this case, which simulations show are correlated with magnetar power \citep{Jerkstrand2017,Dessart2019,OmandJerkstrand2023}. In particular, \citet{OmandJerkstrand2023} show over a large grid of models that oxygen ionization increases as magnetar power increases and ejecta mass decreases.    Finally, factors that reduce ionization, such as ejecta clumping, may occur at varying levels, leading to a diversity of spectral signatures at late times \citep{Jerkstrand2017,Dessart2019}.  Indeed, there may be a continuum from smooth ionized ejecta to highly clumped neutral ejecta.

Our sample shows for the first time that SLSNe indeed exhibit such a continuum, with some events exhibiting a high degree of ionization, while others show no sign of ionization, as inferred from the range of $L_{7300}/L_{6300}$ ratio (highly sensitive to ionized O; Figures \ref{fig:ratiotime} and \ref{fig:specratiopanel}) and the presence of [\ion{O}{3}] $\lambda4363$ and [\ion{O}{3}] $\lambda\lambda4959,5007$ lines, which is correlated with $L_{7300}/L_{6300}$ (Figure \ref{fig:specOIII}).  The [\ion{O}{2}] and [\ion{O}{3}] lines probe lower velocity material than the neutral [\ion{O}{1}] line, indicative of a central ionizing source \citep{ChevalierFransson1992}.  Similar to the [\ion{O}{3}] lines in PS1-14bj and LSQ14an \citep{Lunnan2016,inserra_complexity_2017}, the ratio of the two [\ion{O}{3}] lines is relatively low in our sample, with values of $L_{4959,5007}/L_{4364} \approx 2-3$, indicating electron densities of $n_{\rm e}\sim10^{7}$ cm$^{-3}$ (using the relation from \citealt{OsterbrockFerland2006} and assuming a temperature of $10^{4}$ K).  This is a lower density than that inferred for the region emitting the \ion{O}{1} $\lambda7774$ and [\ion{Ca}{2}] lines (\S\ref{sec:clumping}).      

As discussed in more detail in \S\ref{sec:comps}, such a continuum of ionization levels is not seen in SNe Ic, suggesting it is related to the power source.  On average, we find that slower evolving SLSNe exhibit higher $L_{7300}/L_{6300}$ than faster events further indicating a link with the engine and/or ejecta properties (Figure~\ref{fig:ratio_vs_risedec}).  Moreover, our sample reveals a dominant trend of decreasing $L_{7300}/L_{6300}$ with time characterizing the overall sample (Figure \ref{fig:ratiotime}).  While SLSNe at early phases show the broad continuum of ratios, SLSNe at late phases converge to a tight decline with time, eventually reaching $L_{7300}/L_{6300}\approx 0.2$ at $t/t_{\rm d}\gtrsim8$.  There are three notable outliers (SN\,2017egm, LSQ14an, and PTF10hgi) that not only exhibit high ratios on average, but also an increasing trend with time, which we argue in \S\ref{sec:csm} is likely due to an external source of heating.

Assuming that magnetars power the main population with decreasing $L_{7300}/L_{6300}$, we can explore the engine properties that influence ejecta ionization.  While we find no direct correlation between $L_{7300}/L_{6300}$ and any ejecta or engine parameters individually, we find that late-time energy deposition is the key factor.  In particular, SLSNe with an initially large thermalized spin-down power as they enter the nebular phase exhibit higher ionization on average than events with lower energy deposition at similar phases (Figure~\ref{fig:magpower}).  

This correlation between the thermalized spin-down power and ejecta ionization is further influenced by $M_{\rm ej}$ which affects the magnetar power per unit mass.  Compared to all masses, ejecta ionization in events with $M_{\rm ej} \lesssim 12$ M$_{\odot}$ is more tightly correlated with the thermalized spin-down power; these events have higher energy input per unit mass.  SLSNe with the largest ejecta masses exhibit a weaker correlation.  At later phases of $t/t_{\rm d} \gtrsim4$, when the spin-down power is lower, the oxygen ionization eventually declines due to recombination.  Even so, there remains a statistically significant correlation between the thermalized spin-down power and $L_{7300}/L_{6300}$.  Our sample suggests that ionization breakout may indeed be possible on $\sim$yr timescales, motivating X-ray follow-up of SLSNe with signs of ionization in the early nebular phase to search for breakout emission.

\subsection{The Source of Ionization in Events with Increasing $L_{7300}/L_{6300}$}
\label{sec:csm}

A few SLSNe exhibit increasing ionization with time (while the magnetar energy input declines) pointing to an additional source of ejecta ionization at late times.  A natural explanation is an external source of late-time heating, such as CSM interaction (e.g., heating from the reverse shock propagating through the SN ejecta; \citealt{Fransson1984,ChevalierFransson1994}).  Such a scenario was proposed to explain the [\ion{O}{3}] lines in PS1-14bj \citep{Lunnan2016}.    

Interestingly, the events with increasing ionization all show other signatures of CSM interaction.  SN\,2017egm exhibited prominent late-time light curve bumps \citep{hosseinzadeh_2021,Zhu2023}, possibly due to interaction with shells ejected via the pulsational pair-instability mechanism \citep{Lin2023}, and was identified as an outlier among a large sample of photospheric spectra of SLSNe \citep{Aamer2025}.  PTF10hgi has possible hydrogen and helium absorption lines during the photospheric phase \citep{Quimby2018} and was detected in the radio at late times \citep{Eftekhari2019}.  LSQ14an exhibited light curve undulations \citep{inserra_complexity_2017}, though less prominent than those in SN\,2017egm.

\subsection{Ejecta Clumping}
\label{sec:clumping}

The other line ratios we analyze, $L_{7774}/L_{6300}$ and $L_{7300}/L_{\rm CaIINIR}$, provide constraints on $n_{\rm e}$ and the filling factor, $f$, the fraction of the volume occupied by the ejecta.  \citet{Jerkstrand2017} showed that [\ion{Ca}{2}]/\ion{Ca}{2} NIR $<1$ requires $n_{\rm e} \gtrsim 10^8$ cm$^{-3}$ and therefore low ejecta filling factors.  The low $L_{7300}/L_{\rm CaIINIR}$, combined with the strong, persistent \ion{O}{1} $\lambda7774$, which indicates filling factors of $f\lesssim0.1$ for these high electron densities \citep{Jerkstrand2017,Nicholl2019}, is strong evidence for ejecta clumping in SLSNe.  The magnetar models of \citet{Dessart2019} also show increasing \ion{Ca}{2} NIR and \ion{O}{1} $\lambda7774$ line strength with increasing clumping.  

The low velocities of the \ion{O}{1} $\lambda7774$ and [\ion{Ca}{2}] lines further indicate that they are probing clumping of the innermost material.  In addition, the velocities of the [\ion{O}{2}] and [\ion{O}{3}] lines are similar or only slightly larger than the velocities of \ion{O}{1} $\lambda7774$ and [\ion{Ca}{2}], indicating the presence of both dense recombining clumps and hot ionized cavities in the same physical region of the ejecta (as suspected by \citealt{Jerkstrand2017} and \citealt{Dessart2019} based on modeling of a few SLSNe).  

The inferred clumping may be the result of magnetar-driven hydrodynamic instabilities in the inner ejecta, which have been seen in both 2D and 3D simulations \citep{Chen2016,Chen2020magsim,SuzukiMaeda2021}.  Such clumping may be a ubiquitous effect of a magnetar on SN ejecta.    

\subsection{Differences Between SLSNe and SNe Ic/Ic-BL} 
\label{sec:comps}  

Our large sample of 37 events, which includes the 12 from \citet{Nicholl2019}, allows us to perform a robust comparison with SNe Ic and Ic-BL.  \citet{Fang2019,Fang2022} and \citet{Prentice2022} analyze samples of SNe Ic and SNe Ic-BL as part of larger comparative analyses of SN nebular spectra.  They use a similar Gaussian-fitting procedure to measure the line ratio [\ion{Ca}{2}]/[\ion{O}{1}] (equivalent to $L_{7300}/L_{6300}$).  Unlike the SLSNe in our sample, the SNe Ic/Ic-BL show ratios of $L_{7300}/L_{6300} < 1$ at all times during their evolution.  Even SNe Ic/Ic-BL at relatively early rest-frame phases of $\sim 50-100$ days after peak (corresponding to $t/t_{\rm d} \sim 2-4$) exhibit lower ratios compared to SLSNe, which have ratios of $\approx 0.7-3.8$ during this phase.  It is only at late phases of $t/t_{\rm d} \gtrsim 5$ that SLSNe exhibit similar ratios as SNe Ic/Ic-BL.  

The much larger ratios for SLSNe at early phases represents a significant difference between the two populations, and that SLSNe, unlike SNe Ic/Ic-BL, are able to reach high levels of ejecta ionization due to the presence of a magnetar engine as discussed extensively in \S\ref{sec:ionization}.  The overall higher ionization of SLSNe compared to SNe Ic/Ic-BL is consistent with the higher temperatures of SLSNe in the photospheric phase \citep{Aamer2025}, which has long been suspected to be due to engine powering \citep{Dessart2012,Mazzali2016}.  

SLSNe and SNe Ic/Ic-BL also differ in the ratios $L_{7774}/L_{6300}$ and $L_{7300}/L_{\rm CaIINIR}$.  While all events in our sample show detectable \ion{O}{1} $\lambda7774$, only half of SNe Ic/Ic-BL exhibit this line.  Moreover, the majority of SNe Ic/Ic-BL that show \ion{O}{1} $\lambda7774$ exhibit weaker emission, with ratios of $L_{7774}/L_{6300}\lesssim0.2$ \citep{Taubenberger2009,Nicholl2019}.  Only a few SNe Ic/Ic-BL show strong \ion{O}{1} $\lambda7774$ emission, including SN\,1997dq \citep{Matheson2001} and iPTF15dtg \citep{Taddia2019}.  The SN Ib SN\,2012au also exhibited strong \ion{O}{1} $\lambda7774$ \citep{Milisavljevic2013}, as well as [\ion{O}{2}] and [\ion{O}{3}] lines in a spectrum at 6.2 yr \citep{Milisavljevic2018}.  The similarity of these events to the nebular spectra of SLSNe led to the suggestion that they share the same power source \citep{Milisavljevic2013,Milisavljevic2018}.  Recent magnetar modeling of SN\,2012au showed that clumping is needed to explain the strength of \ion{O}{1} $\lambda7774$ \citep{OmandJerkstrand2023,Dessart2024}.    

Finally, SLSNe and SNe Ic/Ic-BL differ in the evolution of the ratio $L_{7300}/L_{\rm CaIINIR}$. SNe Ic/Ic-BL show increasing ratios of $L_{7300}/L_{\rm CaIINIR}$ from $\approx0.3-0.5$ at $t/t_{\rm d}\sim2-4$ to $\gtrsim2-5$ at $t/t_{\rm d}\gtrsim5$ \citep{Prentice2022}.  The SLSNe in our sample with detections of the \ion{Ca}{2} NIR triplet have ratios of $\lesssim 1.5$, with most at $\approx0.5$.  In contrast with SNe Ic/Ic-BL, SLSNe do not exhibit significant evolution in this ratio.  As discussed in \S\ref{sec:7300}, some SLSNe show a clear [\ion{O}{2}] contribution to the 7300 \AA\ feature, indicating that the actual ratio of [\ion{Ca}{2}]/\ion{Ca}{2} NIR is even lower in these cases.  In summary, the weaker \ion{O}{1} $\lambda7774$ and increasing $L_{7300}/L_{\rm CaIINIR}$ in SNe Ic/Ic-BL compared to SLSNe is likely due to less significant ejecta clumping in SNe Ic/Ic-BL.

\subsection{Lack of Late-time Hydrogen}

No SLSNe in our sample of 37 events exhibit a late-time appearance of hydrogen emission, indicating that interaction with detached neutral hydrogen shells on timescales of $\lesssim$few years after peak is rare in SLSNe ($\lesssim$7 \%).  Assuming a constant velocity since explosion, inferred from the [\ion{O}{1}] emission-line widths, we find that the radii probed at the time of our observations span $\sim2-8\times10^{16}$ cm.  Any hydrogen shells, if present, must therefore be at larger distances and lost by the progenitor at least a decade prior to explosion if ejected in pulsational pair-instability shell ejections (assuming a typical velocity of $\sim1000$ km s$^{-1}$; \citealt{Woosley2007,Woosley2017}) and at least a century before explosion if the hydrogen is lost in a steady $\sim100$ km s$^{-1}$ wind.          

In a previous study, 3 initially hydrogen-poor SLSNe from the Palomar Transient Factory sample were found to exhibit late-time hydrogen emission \citep{Yan2015,Yan2017halpha}.
However, \citet{Yan2017halpha} point out that these events have weaker [\ion{O}{1}] and [\ion{Ca}{2}] lines than typical SLSNe.  While possibly affected by the interaction (e.g., highly ionized by the reverse shock; \citealt{ChevalierFransson1994}), this could indicate these events originate from progenitors with much lower oxygen masses than most SLSNe and therefore may represent a different type of explosion.  This, combined with our non-detection of hydrogen in a large sample, indicates that no representative hydrogen-poor SLSNe have shown late-time hydrogen.

\section{Summary and Conclusions}
\label{sec:conc}

We present and analyze the largest sample to date of nebular-phase spectra of hydrogen-poor SLSNe, with a particular focus on the dominant and relatively isolated spectral features at 6300, 7300, and 7774 \AA.  We fit the line profiles, measure key line ratios, and assess for correlations with light curve timescales and ejecta and engine parameters.  Our large nebular sample further clarifies the typical behavior of SLSNe at these phases, which is consistent with the magnetar picture, but also unveils additional energy sources that can impact the properties of the ejecta at late times.  Our key findings are:

\begin{itemize}
    \item Some SLSNe show clear emission from [\ion{O}{2}] contributing to the 7300 \AA\ feature, along with [\ion{O}{3}] emission, indicating high levels of ejecta ionization in these events.

    \item The sample exhibits a wide range of $L_{7300}/L_{6300}\approx0.7-3.8$ at early phases of $t/t_{\rm d}<4$ before converging to $L_{7300}/L_{6300}\lesssim0.5$ at later phases.  Events with ionized oxygen exhibit the highest ratios, indicating it is ionization, rather than stronger-than-usual [\ion{Ca}{2}], that accounts for the high ratios.  

    \item SLSNe exhibit high, persistent \ion{O}{1} $\lambda7774$ emission and low $L_{7300}/L_{\rm CaIINIR}$ as seen in smaller samples, indicating ejecta clumping at a low velocity coordinate.

    \item The above properties are not seen in normal SNe Ic.

    \item $L_{7300}/L_{6300}$ is correlated with the light curve rise and decline timescales, the first robust correlation between light curve and late-time spectral properties.  

    \item $L_{7300}/L_{6300}$ is not correlated with $P$, $B$, $M_{\rm ej}$, or $E_{\rm KE}$.  However, events with high $L_{7300}/L_{6300}$ have magnetar engines with either long spin-down times or rapid initial spins.

    \item SLSNe which exhibit the dominant decreasing $L_{7300}/L_{6300}$ trend on average have higher ratios with higher instantaneous magnetar energy input.  We find a statistically significant correlation for events with $M_{\rm ej} \lesssim 12$ M$_{\odot}$.  This shows that higher magnetar power leads to higher levels of ejecta ionization, especially for lower mass events which have higher spin-down power per unit mass.  

    \item PTF10hgi, LSQ14an, and SN\,2017egm exhibit the opposite temporal trend of increasing $L_{7300}/L_{6300}$ and thus increasing ionization.  This suggests another source of late-time ionization in these events, likely CSM interaction.

    \item No objects in our sample exhibit late-time H$\alpha$, indicating that interaction with detached neutral H shells on $\lesssim$few year timescales is rare in SLSNe.  
\end{itemize}

Our observations support the theoretical prediction that a magnetar engine can substantially affect the hydrodynamics and ionization state of the inner ejecta of a SN.  More energy deposited at later times leads to greater ionization, which manifests as higher $L_{7300}/L_{6300}$ and stronger [\ion{O}{3}] at the onset of the nebular phase.  Ejecta clumping, driven by fluid instabilities in the inner ejecta due to magnetar energy injection, can create recombining clumps with low filling factor, an effect seen in both 2D and 3D simulations of SLSNe that explains the near universal detection of strong \ion{O}{1} $\lambda$7774 at low velocities.  Moreover, the similar velocities of the [\ion{O}{2}]/[\ion{O}{3}], \ion{O}{1} $\lambda7774$, and [\ion{Ca}{2}] lines show that both dense clumps and hot ionized regions can exist in the same region of the ejecta.  

Importantly, we find that the temporal evolution of $L_{7300}/L_{6300}$ can be used as a diagnostic of the late-time source of ionization, enabling identification of discrepant nebular properties.  Sources like SN\,2017egm and PTF10hgi with persistently high ionization are unlikely to be powered at late times by a magnetar alone.  Our sample highlights the importance of nebular phase observations for testing magnetar predictions and correlating early- and late-time properties.  This is especially important for the upcoming Rubin Observatory when the routine discovery of outliers and unusual events will motivate detailed spectroscopic follow-up to distinguish models.

\begin{acknowledgments}
The Berger Time-Domain Group at Harvard is supported by NSF and NASA grants.  P.K.B.~acknowledges a CIERA Postdoctoral Fellowship, which supported the early stages of this project.  M.N.~is supported by the European Research Council (ERC) under the European Union’s Horizon 2020 research and innovation programme (grant agreement No.~948381).  Some observations reported here were obtained at the MMT Observatory, a joint facility of the Smithsonian Institution and the University of Arizona.  A subset of the MMT observations and all of the Keck Observatory observations were made possible through access supported by Northwestern University and the Center for Interdisciplinary Exploration and Research in Astrophysics (CIERA).  Keck Observatory is a private 501(c)3 non-profit organization operated as a scientific partnership among the California Institute of Technology, the University of California, and the National Aeronautics and Space Administration. The Observatory was made possible by the generous financial support of the W.~M.~Keck Foundation.  The authors wish to recognize and acknowledge the very significant cultural role and reverence that the summit of Maunakea has always had within the Native Hawaiian community. We are most fortunate to have the opportunity to conduct observations from this mountain.  This paper uses data products produced by the OIR Telescope Data Center, supported by the Smithsonian Astrophysical Observatory.  This paper includes data gathered with the 6.5 meter Magellan Telescopes located at Las Campanas Observatory, Chile.  Based in part on observations obtained at the international Gemini Observatory, a program of NSF’s NOIRLab, which is managed by the Association of Universities for Research in Astronomy (AURA) under a cooperative agreement with the National Science Foundation on behalf of the Gemini Observatory partnership: the National Science Foundation (United States), National Research Council (Canada), Agencia Nacional de Investigaci\'{o}n y Desarrollo (Chile), Ministerio de Ciencia, Tecnolog\'{i}a e Innovaci\'{o}n (Argentina), Minist\'{e}rio da Ci\^{e}ncia, Tecnologia, Inova\c{c}\~{o}es e Comunica\c{c}\~{o}es (Brazil), and Korea Astronomy and Space Science Institute (Republic of Korea).  These observations are associated with Gemini programs GN-2018B-FT-111, GN-2019A-Q-233, GS-2019A-Q-234, GN-2021A-Q-204, GS-2022A-Q-207, GN-2022A-Q-204, GS-2022B-Q-104, GS-2022B-Q-701, GS-2023A-Q-204, GN-2023B-Q-241, GN-2024B-Q-207, GS-2024B-Q-106, GN-2025A-Q-108, and GS-2025A-Q-108. 
\end{acknowledgments} 

\software{Astropy \citep{astropy:2013,astropy:2018,astropy:2022}, IRAF \citep{Tody1986,Tody1993}, Pypeit \citep{Pypeit}, redspec \citep{redspec2024}}

\facilities{MMT, Magellan, Keck, Gemini}

\bibliography{references}
\bibliographystyle{aasjournalv7}

\end{document}